\RequirePackage{fix-cm}
\documentclass[smallcondensed,notitlepage]{svjour3}
\usepackage{amsmath}
\smartqed

\usepackage[T1]{fontenc}
\usepackage[utf8]{inputenc}
\usepackage{newtxtext,newtxmath}

\usepackage{graphicx}
\usepackage{amsmath,mathtools}        
\usepackage{booktabs,multirow,array}
\usepackage{subcaption}                
\usepackage{natbib}
\usepackage{hyperref}
\usepackage{float} 
\usepackage{rotating}
\usepackage{tcolorbox}

\newcommand{\SFL}{\mathrm{S}\!\downarrow\!\mathrm{F}\!\uparrow\!\mathrm{L}}
\newcommand{\LFS}{\mathrm{L}\!\downarrow\!\mathrm{F}\!\uparrow\!\mathrm{S}}

\graphicspath{{figs/}{figures/}{images/}}

\journalname{Empirical Software Engineering}

\title{Understanding Chain-of-Thought Effectiveness in Code Generation: An Empirical and Information-Theoretic Analysis}

\titlerunning{CoT Methods for Code Generation}

\author{Naizhu~Jin \and Zhong~Li\thanks{* Corresponding author} \and Guang~Yang \and Tian~Zhang \and Qingkai~Zeng}
\authorrunning{Jin et al.}

\institute{
Naizhu~Jin \at State Key Laboratory for Novel Software Technology, Nanjing University, Nanjing, China \\
\email{jnz@smail.nju.edu.cn}
\and
Zhong~Li* \at State Key Laboratory for Novel Software Technology, Nanjing University, Nanjing, China \\
\email{lizhong@nju.edu.cn}\
\and
Guang~Yang \at The State Key Laboratory of Blockchain and Data Security, Zhejiang University‌, Hangzhou, China \\
\email{novelyg@outlook.com}\
\and
Tian~Zhang \at State Key Laboratory for Novel Software Technology, Nanjing University, Nanjing, China \\
\email{ztluck@nju.edu.cn}
\and
Qingkai~Zeng \at State Key Laboratory for Novel Software Technology, Nanjing University, Nanjing, China \\
\email{zqk@nju.edu.cn}
}

\date{Received: date / Accepted: date}

\begin{document}
\maketitle

\begin{abstract}
Large language models (LLMs) achieve strong performance on code generation, but the mechanisms by which Chain-of-Thought (CoT) prompting helps remain unclear.
We present a systematic empirical and information-theoretic study of CoT effectiveness in neural code generation, evaluating five paradigms (Zero-Shot, Zero-Shot CoT, Self-Planning, Structured CoT, Reasoning-CoT) across six Python benchmarks, a multilingual benchmark with 12 programming languages, and six models from 7B to 480B parameters, using conditional mutual information $I(Y;C|X)$ as a conceptual lens.
Our results show that externally guided CoT consistently outperforms direct generation, with structured methods improving Pass@1 by 5--12\% on average while using substantially fewer tokens than reflective reasoning, and that CoT benefits depend on language type systems and model capacity.
We further find that reasoning \emph{quality} is critical: high-quality structured CoT from strong generators yields significantly higher accuracy than lightweight alternatives with the same template, whereas naive Zero-Shot CoT can even degrade performance.
These findings provide practical guidance for choosing CoT strategies based on model capacity, language characteristics, and task complexity.
    \end{abstract}


\section{Introduction}
\label{sec:intro}
Large language models (LLMs) have fundamentally transformed automated code generation, enabling the translation of natural language requirements into executable programs with unprecedented accuracy~\cite{nijkamp2022codegen,fried2022incoder,vaithilingam2022expectation,roziere2023code,wang2023natural}. 
Among various prompting techniques, CoT reasoning~\cite{wei2022chain,kojima2022large,ma2023bridging} has emerged as a particularly effective approach, encouraging models to generate explicit intermediate reasoning steps before producing final code, thereby improving both correctness and interpretability~\cite{lyu2023faithful,jie2024interpretable,stechly2024chain}.

Building upon the success of standard CoT (i.e., using the simple prompt ``Let's think step by step'' to elicit reasoning), numerous variants have been proposed to enhance reasoning structure and control in code generation~\cite{yao2023tree,bi2024program,besta2024graph}.
These methods can be broadly categorized into three paradigms based on their reasoning mechanisms:
\textbf{(1) Prompt-elicited reasoning}, which triggers spontaneous step-by-step thinking through simple prompts without explicit structure;
\textbf{(2) Structure-constrained reasoning}, which guides the model through predefined templates or hierarchical planning to ensure consistency and interpretability;
and \textbf{(3) Reflective reasoning}, which incorporates iterative self-correction and deep exploration into multi-step reasoning processes.
These paradigms collectively represent a spectrum of reasoning strategies from lightweight prompt-based elicitation to heavyweight reflective exploration, each with distinct trade-offs in expressiveness, efficiency, and reliability~\cite{turpin2023language}.

Despite growing interest in CoT-based code generation, fundamental questions remain unanswered.
\textbf{(1) Lack of systematic cross-paradigm evaluation:} While individual CoT methods report performance gains on specific benchmarks, systematic comparisons across diverse paradigms, models, datasets, and programming languages are lacking. It remains unclear when and why different reasoning structures succeed or fail, and whether CoT benefits generalize beyond Python to languages with different type systems and syntactic characteristics~\cite{fu2023chain}.
\textbf{(2) Unexplored model capacity effects:} Smaller models may require explicit reasoning scaffolding, while larger models might internalize reasoning patterns during pretraining. However, no systematic analysis quantifies how model scale affects CoT responsiveness, nor characterizes the asymmetric success and failure patterns that emerge when identical reasoning is applied to models of different capacities~\cite{diao2024active}.
\textbf{(3) Uncharacterized quality-effectiveness relationship:} Existing work focuses on paradigm-level comparisons without examining \textit{what information} reasoning chains provide and \textit{how} reasoning quality influences generation accuracy. The trade-offs between reasoning depth, computational cost, and downstream reliability remain poorly understood, limiting practitioners' ability to select appropriate CoT strategies for real-world deployments~\cite{zhao2025trade}.

To address these gaps, we formulate four research questions that systematically examine CoT effectiveness from complementary perspectives:

\noindent\textbf{RQ1 (Performance-Efficiency Trade-offs):} How do different CoT paradigms compare in terms of generation accuracy and token cost across Python benchmarks of varying difficulty?

\noindent\textbf{RQ2 (Cross-Language Generalization):} How do CoT paradigms perform across 12 programming languages with diverse type systems and syntactic characteristics?

\noindent\textbf{RQ3 (Capacity-Dependent Patterns):} Why does the same CoT paradigm succeed on larger models but fail on smaller models within the same family?

\noindent\textbf{RQ4 (Quality-Effectiveness Relationship):} How does reasoning quality affect code generation accuracy when using the same paradigm structure?

To answer these questions, we first develop an information-theoretic framework that formalizes CoT effectiveness through conditional mutual information $I(Y;C|X)$, which quantifies how reasoning chains reduce generation uncertainty.
This framework yields four testable hypotheses: (H1) structured paradigms achieve higher information density $I(Y;C|X)/L$ than reflective paradigms; (H2) CoT benefits vary across languages, with statically typed languages showing different responsiveness than dynamically typed ones; (H3) smaller models fail to extract information from reasoning chains when complexity exceeds their capacity; and (H4) reasoning quality, not just paradigm structure, determines CoT effectiveness.

We validate these hypotheses through extensive experiments across six models (Qwen2.5-Coder-7B/32B, Qwen3-Coder-30B/480B, GPT-3.5-Turbo, GPT-5), six Python benchmarks, and 12 programming languages.
Our key findings are:
\begin{itemize}
    \item \textbf{RQ1:} Structured paradigms (Self-Planning, SCoT) achieve 85--95\% of Reasoning-CoT's accuracy while using only $\sim$10\% of its tokens. Naive Zero-Shot CoT often degrades performance due to reasoning hallucinations.
    
    \item \textbf{RQ2:} CoT benefits generalize across 12 languages, with structured paradigms yielding larger gains in statically typed languages (+7\%) and reflective reasoning providing more balanced improvements in dynamically typed languages (+6\%).
    
    \item \textbf{RQ3:} Larger models succeed in $\sim$67\% of asymmetric cases, while smaller models exhibit systematic failures in type/boundary handling and reasoning-execution alignment that larger models resolve.
    
    \item \textbf{RQ4:} High-quality reasoning (GPT-5-Mini) outperforms lightweight alternatives (7B model) by 7.5\% Pass@1 under identical templates, and low-quality CoT can even harm performance below Zero-Shot baselines.
\end{itemize}

\noindent The remainder of this paper is organized as follows. Section~\ref{sec:background} reviews neural code generation and Chain-of-Thought prompting methods and positions our work within existing comparative studies. Section~\ref{sec:theory} introduces the information-theoretic framework for characterizing CoT effectiveness, and Section~\ref{sec:setup} details the research questions, CoT paradigms, datasets, models, and evaluation metrics. 
The subsequent sections present empirical results for RQ1--RQ4, followed by a discussion of threats to validity and the conclusion and future directions in Section~\ref{sec:conclusion}.

\section{Background and Related Work}
\label{sec:background}

\subsection{Neural Code Generation}

Large language models (LLMs) have transformed code generation from rule-based synthesis to neural end-to-end learning~\cite{feng2020codebert,li2022competition}.
Given a natural language description $X$ and target code $Y$, a code generation model $M_{\theta}$ learns to maximize the conditional probability $P_\theta(Y|X)$ through autoregressive factorization:
\begin{equation}
P_\theta(Y|X) = \prod_{t=1}^{|Y|} P_\theta(y_t | X, y_{<t}),
\end{equation}
where $y_t$ denotes the $t$-th token and $y_{<t}$ represents all preceding tokens.

Early approaches relied on sequence-to-sequence models~\cite{iyer2016summarizing,yin2017syntactic}, while recent pre-trained code models such as CodeBERT~\cite{feng2020codebert}, CodeT5~\cite{wang2021codet5}, Codex~\cite{chen2021evaluating}, and StarCoder~\cite{lozhkov2024starcoder} leverage large-scale pretraining to achieve strong zero-shot and few-shot performance across multiple programming languages.
Despite these advances, LLMs struggle with complex tasks requiring multi-step reasoning, algorithmic planning, and structured problem decomposition~\cite{yang2024important,jiang2024survey}, motivating the need for explicit reasoning mechanisms.

\subsection{Chain-of-Thought Prompting Methods}

CoT prompting~\cite{wei2022chain} addresses these limitations by encouraging models to generate intermediate reasoning steps before producing final outputs.
In code generation, CoT extends the autoregressive process to condition on explicit reasoning chains $C$:
\begin{equation}
P(Y|X) \approx P_{\text{cot}}(C|X) \cdot P_{\text{code}}(Y|X, C),
\end{equation}
where $C$ represents step-by-step reasoning that decomposes the problem and guides code synthesis.

Building on this foundation, numerous CoT variants have been proposed.
\textbf{Zero-shot CoT}~\cite{kojima2022large} elicits reasoning through simple prompts (e.g., ``Let's think step by step''), enabling spontaneous reasoning without exemplars.
\textbf{Few-shot CoT}~\cite{wei2022chain} provides demonstration examples with explicit reasoning traces to guide model behavior.
\textbf{Self-Planning CoT}~\cite{jiang2024self} introduces hierarchical task decomposition, where models first generate high-level plans before implementing code.
\textbf{Structured CoT (SCoT)}~\cite{li2025structured} constrains reasoning within predefined templates (e.g., problem analysis $\rightarrow$ algorithm design $\rightarrow$ implementation), ensuring consistency and interpretability.
\textbf{Multilingual SCoT (MSCoT)}~\cite{jin2025mscot} extends structured reasoning to cross-language scenarios with lightweight slot-based prompts.
\textbf{Reasoning-CoT} integrates reflective self-correction and verification into multi-step reasoning.

Beyond prompting methods, researchers have explored CoT dataset construction~\cite{yang2024chain} and compression techniques~\cite{miao2024chain} to improve efficiency.
Liu et al.~\cite{liurevisiting} investigated whether LLMs internalize reasoning before code generation, revealing that reasoning benefits depend on model architecture and scale.
However, these studies focus on individual paradigms or isolated model-dataset pairs, lacking systematic cross-paradigm comparison.

\subsection{Comparative Studies and Research Gaps}

While individual CoT methods demonstrate empirical improvements, comprehensive comparative analyses remain scarce.
Most prior work evaluates single paradigms on specific benchmarks~\cite{yang2024chain,jin2025mscot}, reporting aggregate performance gains without systematic cross-paradigm comparison.
Recent studies have begun examining CoT behavior in code generation: Liu et al.~\cite{liurevisiting} investigated whether LLMs internalize reasoning before code generation, revealing that reasoning benefits depend on model architecture and scale; Yang et al.~\cite{yang2024chain} explored CoT dataset construction for code synthesis.
However, these efforts remain fragmented, focusing on isolated paradigm-dataset pairs rather than providing unified empirical and theoretical understanding.

Our work differs from prior studies in three aspects: (1) we systematically compare multiple CoT paradigms across diverse models, datasets, and programming languages; (2) we introduce an information-theoretic lens to characterize \textit{why} and \textit{when} CoT succeeds; and (3) we provide controlled analyses of model capacity effects and reasoning quality, offering actionable guidance for practitioners.

\section{Theoretical Framework}
\label{sec:theory}

To formalize CoT effectiveness beyond empirical observation, we introduce an information-theoretic perspective that quantifies how reasoning chains reduce generation uncertainty.
This framework provides both a theoretical lens for interpreting our empirical findings and a principled basis for comparing different CoT paradigms.

\subsection{Information-Theoretic View of CoT}

Given the probabilistic formulation $P(Y|X) \approx P_{\text{cot}}(C|X) \cdot P_{\text{code}}(Y|X, C)$, the reasoning chain $C$ functions as an auxiliary information channel transmitting task-relevant signals from input $X$ to output $Y$.
The effectiveness of this channel can be quantified by conditional mutual information:
\begin{equation}
I(Y;C|X) = H(Y|X) - H(Y|X,C),
\label{eq:mutual_info}
\end{equation}
where $H(\cdot)$ denotes Shannon entropy.
This metric measures how much additional information $C$ provides about the correct code $Y$ beyond the input $X$.

\textbf{Intuition:} $H(Y|X)$ represents the uncertainty in code generation given only the problem description, while $H(Y|X,C)$ represents the remaining uncertainty after observing the reasoning chain.
Under an optimal decoder, the improvement in expected log-likelihood equals $I(Y;C|X)$, indicating that every bit of useful reasoning information directly enhances generation accuracy.
Conversely, when $C$ introduces irrelevant or hallucinated reasoning, it may fail to reduce $H(Y|X,C)$ or even increase uncertainty, explaining inconsistent CoT benefits observed in practice~\cite{zhao2025chain}.

    
    
    

\subsection{Information Capacity and Efficiency Trade-offs}

The information capacity of $C$ is bounded by its token length $L$ and vocabulary size $V$:
\begin{equation}
I(Y;C|X) \leq L \log_2 V.
\label{eq:capacity_bound}
\end{equation}
This bound implies that longer reasoning chains can encode more information but also risk introducing noise and redundancy.

Different CoT paradigms navigate this information-efficiency frontier differently:
\textbf{Structured paradigms} (e.g., Self-Planning, SCoT) use short ($L \sim 200$--$700$ tokens), template-constrained reasoning to preserve essential information with low entropy, optimizing for efficiency and consistency.
\textbf{Reflective paradigms} (e.g., Reasoning-CoT) employ longer ($L \sim 2000$--$7000$ tokens), free-form reasoning to explore deeper problem decomposition, optimizing for expressiveness at higher computational cost.

Based on this framework, we derive four testable hypotheses (H1--H4) as outlined in Section~\ref{sec:intro}, which guide our empirical analysis in subsequent sections.
In our experiments, we use observable proxies: Pass@1 with token cost for information density (RQ1), cross-language performance for language-dependent informativeness (RQ2), asymmetric success/failure rates for capacity effects (RQ3), and quality-induced performance gaps for reasoning informativeness (RQ4).

\section{Experimental Setup}
\label{sec:setup}

This section describes the experimental design for systematically evaluating CoT effectiveness in neural code generation. Our study addresses four research questions that examine CoT paradigms from complementary perspectives: performance-cost trade-offs, cross-language generalization, capacity-dependent responsiveness, and quality-dependent effectiveness.

\subsection{Research Questions and Experimental Design}

As outlined in Section~\ref{sec:intro}, our study addresses four research questions examining CoT effectiveness from complementary perspectives. Here we detail the experimental design for each:

\textbf{RQ1} compares five paradigms (Zero-Shot, Zero-Shot CoT, Self-Planning, SCoT, Reasoning-CoT) across six Python benchmarks spanning entry-level (MBPP) to complex (BigCodeBench) difficulty, measuring both Pass@1 and token consumption.

\textbf{RQ2} evaluates cross-language generalization on HumanEval-XL (12 languages), comparing statically typed (Java, C++, Go, etc.) and dynamically typed (JavaScript, Ruby, PHP, etc.) languages.

\textbf{RQ3} uses controlled model pairs (GPT-3.5 vs. GPT-5, Qwen2.5-7B vs. 32B, Qwen3-30B vs. 480B) with identical SCoT prompts to identify capacity-dependent asymmetric outcomes.

\textbf{RQ4} compares high-quality SCoT (GPT-5-Mini) against lightweight MSCoT (7B model) under identical template structures across all downstream models.

\subsection{CoT Paradigms}

We evaluate six representative CoT paradigms spanning the spectrum from direct generation to structured reasoning. Table~\ref{tab:cot_paradigms} summarizes their characteristics and generation methods.

\begin{table}[h]
\centering
\caption{Overview of CoT paradigms and their generation methods.}
\label{tab:cot_paradigms}
\small
\begin{tabular}{llll}
\toprule
\textbf{Paradigm} & \textbf{Type} & \textbf{Generator} & \textbf{Implementation} \\
\midrule
Zero-Shot & Baseline & -- & Direct code generation \\
Zero-Shot CoT & Spontaneous & -- & Prompt: ``Let's think step by step'' \\
Self-Planning & Hierarchical & GPT-5-Mini & 3-shot ICL: decomposition $\rightarrow$ plan $\rightarrow$ code \\
SCoT & Structured & GPT-5-Mini & 3-shot ICL: template-constrained slots \\
Reasoning-CoT & Reflective & DeepSeek-R1 & Zero-shot deep reasoning with self-correction \\
\bottomrule
\end{tabular}
\end{table}

\textbf{Zero-Shot} serves as the baseline, generating code directly from problem descriptions without intermediate reasoning steps.

\textbf{Zero-Shot CoT} elicits spontaneous reasoning by appending ``Let's think step by step'' to problem descriptions, encouraging the model to produce reasoning chains before code without requiring exemplars.

\textbf{Self-Planning} uses GPT-5-Mini with 3-shot in-context learning to generate hierarchical reasoning. Given exemplars demonstrating task decomposition, the model first breaks problems into subtasks, outlines their execution order, then generates code following the plan. This two-stage structure emphasizes top-down problem solving.

\textbf{SCoT (Structured CoT)} also employs GPT-5-Mini with 3-shot in-context learning, but constrains reasoning within predefined template slots (e.g., sequence/branch/loop structures). The fixed structure ensures consistency and filters irrelevant reasoning, optimizing information density.


\textbf{Reasoning-CoT} leverages DeepSeek-R1, an open-source reasoning-oriented model trained with reinforcement learning. Without requiring exemplars, DeepSeek-R1 generates long-form reflective reasoning chains featuring explicit problem decomposition, multiple solution explorations, and self-correction mechanisms. This paradigm prioritizes reasoning depth and interpretability over efficiency.


\subsection{Datasets}

We evaluate CoT paradigms on seven benchmarks covering diverse difficulty levels, reasoning complexity, and programming languages. Six Python benchmarks (RQ1) assess performance across varying task complexity, while one multilingual benchmark (RQ2) evaluates cross-language generalization. 

\subsubsection{Python Benchmarks (RQ1)}

\textbf{MBPP}~\cite{austin2021program} contains 974 entry-level Python tasks written by crowdworkers, with an average of 3 test cases per problem. It evaluates fundamental code synthesis ability on simple function-level tasks.

\textbf{HumanEval}~\cite{chen2021evaluating} comprises 164 hand-crafted Python problems with function signatures, docstrings, and 7--8 test cases each. It serves as the standard benchmark for functional correctness evaluation in neural code generation.

\textbf{OpenEval}~\cite{yang2024chain} includes 178 reasoning-oriented problems emphasizing multi-step logical planning and natural language understanding during code synthesis.

\textbf{MHPP}~\cite{dai2024mhpp} consists of 210 manually constructed problems across seven challenge categories, with 14 test cases per problem on average. Problems feature 167.6 words per description and 14.9 lines per reference solution, reflecting greater algorithmic complexity than MBPP and HumanEval.

\textbf{CodeHarmony}~\cite{wei2023towards} provides a test subset of 153 samples (from 16,153 total) synthesized by LLMs from curated open-source corpora. It evaluates reasoning consistency and semantic alignment between docstrings and generated code.

\textbf{BigCodeBench}~\cite{zhuo2024bigcodebench} comprises approximately 1,140 Python tasks with 33.5 lines of code and 5--6 test cases per problem on average. It emphasizes complex instruction following, diverse function calling, and real-world compositional reasoning.

\subsubsection{Multilingual Benchmark (RQ2)}

\textbf{HumanEval-XL}~\cite{peng2024humaneval} extends HumanEval to 12 programming languages: Python, Java, JavaScript, C++, C\#, PHP, Ruby, Go, Rust, Swift, Kotlin, and TypeScript. With 164 problems per language (1,968 total), it enables systematic evaluation of CoT effectiveness across statically typed languages (Java, C++, C\#, TypeScript, Go, Rust, Swift, Kotlin) and dynamically typed languages (Python, JavaScript, Ruby, PHP).



\subsection{\textbf{Code Generation Models.}}  

We evaluate six code generation models spanning 7B to 480B parameters, covering both proprietary and open-source architectures. All models are accessed through official APIs or open-source releases as of mid-2025.

\textbf{Qwen2.5-Coder-7B\footnote{\url{https://huggingface.co/Qwen/Qwen2.5-Coder-7B-Instruct}}/32B-Instruct\footnote{\url{https://huggingface.co/Qwen/Qwen2.5-Coder-32B-Instruct}} (Alibaba DAMO)} are open-source dense architectures optimized for code generation. The 7B and 32B variants enable controlled capacity-dependent analysis, examining how model scale affects CoT responsiveness within the same model family.

\textbf{Qwen3-Coder-30B-A3B\footnote{\url{https://huggingface.co/Qwen/Qwen3-Coder-30B-A3B-Instruct}}/480B-A35B-Instruct\footnote{\url{https://huggingface.co/Qwen/Qwen3-Coder-480B-A35B-Instruct}} (Alibaba DAMO)} are open-source mixture-of-experts (MoE) architectures with 30B and 480B total parameters, activating 3B and 35B parameters per token respectively. These large-scale models evaluate CoT effectiveness at higher capacity levels.

\textbf{GPT-3.5-Turbo\footnote{\url{https://platform.openai.com/docs/models/gpt-3.5-turbo}} 
and GPT-5\footnote{\url{https://platform.openai.com/docs/models/gpt-5}}
 (OpenAI)} are proprietary models serving as commercial baselines. GPT-3.5-Turbo represents smaller-scale proprietary models, while GPT-5 provides a high-capacity frontier baseline for comparison with open-source alternatives.

\begin{table}[ht]
\centering
\caption{Overview of code generation models evaluated in this study.}
\label{tab:models}
\small
\begin{tabular}{lccc}
\toprule
\textbf{Model} & \textbf{Parameters} & \textbf{Architecture} & \textbf{Access} \\
\midrule
Qwen2.5-Coder-7B-Instruct & 7B & Dense & Open-source \\
Qwen2.5-Coder-32B-Instruct & 32B & Dense & Open-source \\
Qwen3-Coder-30B-A3B-Instruct & 30B (3B active) & MoE & Open-source \\
Qwen3-Coder-480B-A35B-Instruct & 480B (35B active) & MoE & Open-source \\
GPT-3.5-Turbo & -- & Dense & API \\
GPT-5 & -- & Dense & API \\
\bottomrule
\end{tabular}
\end{table}

This selection covers diverse architectures (dense vs. MoE), parameter scales (7B to 480B), and access modes (open-source vs. proprietary), enabling comprehensive evaluation of CoT paradigms across model capacities and families. The key characteristics of all evaluated models are summarized in Table \ref{tab:models}.

\subsection{Evaluation Metric}

We use \textbf{Pass@1} as the primary metric, measuring the proportion of generated programs that pass all test cases on the first attempt:
\begin{equation}
\text{pass@1}=\frac{1}{N}\sum_{i=1}^{N}\mathbb{1}\{\text{code}_i \text{ passes all tests}\},
\end{equation}
where $N$ is the total number of problems and $\mathbb{1}\{\cdot\}$ is the indicator function. This metric directly reflects functional correctness, the primary criterion for evaluating code generation quality.

\section{Results Analysis}\label{sec:results}

\subsection{RQ1: Performance and efficiency trade-offs across Python benchmarks}  

\begin{table}[t]
    \centering
    \caption{RQ1 results: Pass@1 (\%) comparison across Python benchmarks of varying difficulty. Best result per model in \textbf{bold}.}
    \label{tab:rq1_results}
    \resizebox{\textwidth}{!}{
    \begin{tabular}{l|ccccc|ccccc}
    \toprule
    & \multicolumn{5}{c|}{\textbf{Qwen2.5-Coder-7B-Instruct}} & \multicolumn{5}{c}{\textbf{Qwen2.5-Coder-32B-Instruct}} \\
    \cmidrule(lr){2-6} \cmidrule(lr){7-11}
    \textbf{Dataset} & Zero & CoT & Self-P & SCoT & Reas & Zero & CoT & Self-P & SCoT & Reas \\
    \midrule
    BigCodeBench & 18.24 & 14.86 & 20.27 & 16.89 & 28.38 & 29.73 & 31.08 & 29.05 & 22.97 & 29.73 \\
    CodeHarmony & 60.13 & 57.52 & 59.48 & 60.13 & 61.44 & 67.32 & 66.67 & 65.36 & 64.71 & 62.75 \\
    HumanEval & 78.05 & 78.66 & 82.32 & 85.98 & 85.98 & 89.63 & 88.41 & 92.68 & 90.85 & 93.90 \\
    MBPP & 57.20 & 50.60 & 52.60 & 52.80 & 58.40 & 57.00 & 56.40 & 54.00 & 54.60 & 60.00 \\
    MHPP & 8.57 & 21.90 & 37.14 & 38.10 & 39.05 & 47.62 & 46.19 & 61.43 & 57.14 & 57.14 \\
    OpenEval & 39.33 & 44.38 & 47.19 & 45.51 & 46.07 & 49.44 & 41.57 & 48.31 & 47.19 & 50.00 \\
    \midrule
    \textbf{Average} & 44.65 & 43.59 & 49.83 & 49.90 & \textbf{53.22} & 56.79 & 55.05 & 58.47 & 56.24 & \textbf{58.92} \\
    \toprule
    & \multicolumn{5}{c|}{\textbf{Qwen3-Coder-30B-A3B-Instruct}} & \multicolumn{5}{c}{\textbf{Qwen3-Coder-480B-A35B-Instruct}} \\
    \cmidrule(lr){2-6} \cmidrule(lr){7-11}
    \textbf{Dataset} & Zero & CoT & Self-P & SCoT & Reas & Zero & CoT & Self-P & SCoT & Reas \\
    \midrule
    BigCodeBench & 31.76 & 14.19 & 29.05 & 22.30 & 30.41 & 31.76 & 14.19 & 30.41 & 25.68 & 34.46 \\
    CodeHarmony & 67.97 & 68.63 & 69.28 & 67.32 & 64.71 & 69.93 & 69.28 & 68.63 & 69.28 & 62.75 \\
    HumanEval & 92.07 & 85.98 & 94.51 & 95.73 & 96.34 & 93.90 & 85.98 & 96.95 & 96.34 & 93.90 \\
    MBPP & 53.20 & 57.80 & 58.20 & 58.20 & 62.40 & 61.40 & 60.00 & 58.00 & 57.40 & 60.80 \\
    MHPP & 45.24 & 39.05 & 57.14 & 56.19 & 54.29 & 51.43 & 46.19 & 65.24 & 62.86 & 57.62 \\
    OpenEval & 51.12 & 48.31 & 51.69 & 49.44 & 47.75 & 51.69 & 50.00 & 50.56 & 48.88 & 50.00 \\
    \midrule
    \textbf{Average} & 56.89 & 52.33 & \textbf{59.98} & 58.20 & 59.32 & 60.02 & 54.27 & \textbf{61.63} & 60.07 & 59.92 \\
    \toprule
    & \multicolumn{5}{c|}{\textbf{GPT-3.5-Turbo}} & \multicolumn{5}{c}{\textbf{GPT-5}} \\
    \cmidrule(lr){2-6} \cmidrule(lr){7-11}
    \textbf{Dataset} & Zero & CoT & Self-P & SCoT & Reas & Zero & CoT & Self-P & SCoT & Reas \\
    \midrule
    BigCodeBench & 16.22 & 19.59 & 22.97 & 22.97 & 29.73 & 29.05 & 31.08 & 29.73 & 24.32 & 32.43 \\
    CodeHarmony & 70.59 & 69.93 & 64.71 & 63.40 & 66.01 & 67.97 & 62.09 & 65.36 & 66.67 & 67.32 \\
    HumanEval & 68.29 & 68.29 & 89.02 & 90.24 & 95.73 & 95.73 & 96.34 & 96.34 & 94.51 & 95.73 \\
    MBPP & 52.80 & 51.00 & 53.80 & 54.60 & 59.80 & 58.00 & 55.60 & 55.80 & 57.40 & 58.60 \\
    MHPP & 23.81 & 25.71 & 52.38 & 50.95 & 49.05 & 57.14 & 61.43 & 68.10 & 65.71 & 64.76 \\
    OpenEval & 42.70 & 44.38 & 48.31 & 48.31 & 50.00 & 51.12 & 48.31 & 51.12 & 51.69 & 48.88 \\
    \midrule
    \textbf{Average} & 46.48 & 45.74 & 55.20 & 55.08 & \textbf{58.39} & 59.84 & 59.14 & 61.08 & 60.05 & \textbf{61.29} \\
    \bottomrule
    \end{tabular}
    }
\end{table}

\begin{figure}[t]
    \centering
    \includegraphics[width=1\textwidth]{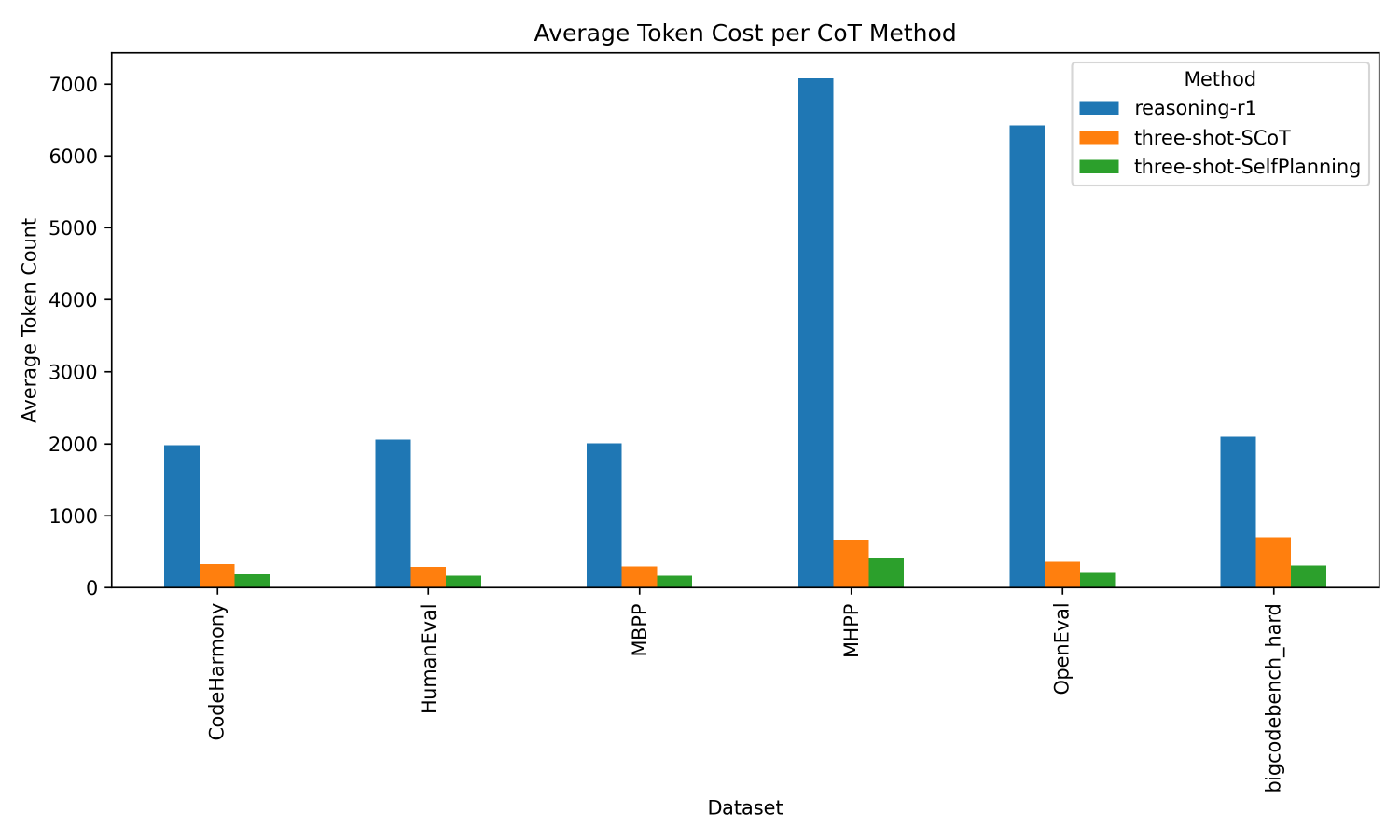}
    \caption{Token Cost Comparison.}
    \label{fig:token_cost_comparison}
\end{figure}

RQ1 examines the fundamental trade-off between reasoning depth and computational efficiency by evaluating five CoT paradigms across six Python benchmarks of varying difficulty and real-world complexity. 
We analyze both Pass@1 accuracy and token consumption to identify which paradigms achieve favorable performance-cost ratios.

\noindent\textbf{Key findings.}
Across six Python benchmarks and six models, externally guided CoT paradigms (Self-Planning, SCoT, Reasoning-CoT) consistently outperform direct generation, with average Pass@1 gains of roughly 5--12\%.
Structured paradigms (Self-Planning, SCoT) achieve 85--95\% of Reasoning-CoT's accuracy while using about one-tenth of its tokens, indicating higher information density.
CoT benefits grow with task difficulty and are largest for smaller models, whereas larger models show diminishing returns from external reasoning.

\subsubsection{Overall Performance}  
Table~\ref{tab:rq1_results} reveals that external reasoning guidance substantially outperforms baseline approaches across all models. 
Averaging across all six models, Reasoning-CoT achieves the highest mean accuracy (58.86\%), followed closely by Self-Planning (57.60\%) and SCoT (56.65\%), while Zero-Shot CoT (52.03\%) performs worse than even Zero-Shot baseline (54.10\%). 
This indicates that \textit{unguided self-generated reasoning often degrades performance}, likely due to reasoning hallucinations where models produce logically coherent but functionally incorrect intermediate steps.

In contrast, externally guided paradigms (Self-Planning, SCoT, Reasoning-CoT) consistently improve over baseline by 5--12\% on average, with gains most pronounced in smaller models (GPT-3.5: +12.65\% for Reasoning-CoT) and diminishing in larger models (GPT-5: +1.45\% for Self-Planning). 
This capacity-modulated responsiveness suggests that external reasoning compensates for limited intrinsic reasoning in smaller models but becomes redundant when models already possess strong internal reasoning circuits.

\subsubsection{Token Cost Analysis}  
Figure~\ref{fig:token_cost_comparison} quantifies the computational cost of each paradigm. Reasoning-CoT consumes 2,000--7,000 tokens per problem, compared to 200--700 tokens for Self-Planning and SCoT. 
Despite this 10$\times$ cost difference, Reasoning-CoT's accuracy advantage is marginal: only +1.26\% over Self-Planning and +2.21\% over SCoT on average.

Thus, structured paradigms (Self-Planning, SCoT) achieve better efficiency than Reasoning-CoT, which is consistent with hypothesis \textbf{H1} that structured reasoning maximizes $I(Y;C|X)/L$; here we use Pass@1 together with token cost as a practical proxy for information density.

\subsubsection{Task Complexity Dependency.}  
CoT effectiveness exhibits strong task-dependent patterns. 
On entry-level benchmarks (MBPP), reasoning methods yield minimal gains (+1--3\%) and sometimes underperform Zero-Shot baselines, indicating that simple function synthesis does not benefit from explicit decomposition. 
In contrast, complex reasoning-intensive benchmarks show substantial improvements:

\begin{itemize}
    \item \textbf{MHPP}: Structured paradigms improve accuracy by 15--30\% over Zero-Shot across all models, with GPT-3.5 gaining +28.57\% (Self-Planning). This substantial improvement reflects CoT's ability to decompose hierarchical task dependencies.
    
    \item \textbf{BigCodeBench}: Reasoning-CoT achieves consistent gains (+8--13\%) across models, handling compositional reasoning and diverse API interactions more effectively than structured paradigms.
    
    \item \textbf{HumanEval}: Structured paradigms (Self-Planning, SCoT) yield +12--22\% improvements in smaller models (GPT-3.5, Qwen2.5-7B) but only +1--3\% in larger models (GPT-5, Qwen3-480B), confirming capacity-modulated responsiveness.
\end{itemize}

This difficulty-dependent scaling validates that CoT paradigms primarily reduce uncertainty $H(Y|X,C)$ when baseline uncertainty $H(Y|X)$ is high (complex tasks), but provide limited information gain when tasks are straightforward.

\subsubsection{Model Capacity and CoT Responsiveness}  
Within-family comparisons reveal systematic capacity effects:

\begin{itemize}
    \item \textbf{Qwen2.5-Coder (7B $\rightarrow$ 32B):} Scaling from 7B to 32B improves Zero-Shot baseline by +12.14\%, while reducing CoT benefits: Reasoning-CoT's advantage drops from +8.57\% to +2.13\%. This indicates that larger models internalize reasoning patterns during pretraining, diminishing the marginal utility of external guidance.

    \item \textbf{Qwen3-Coder (30B $\rightarrow$ 480B):} Despite a 16$\times$ parameter increase, performance gains are modest (+3.13\% average), and Self-Planning remains the most effective paradigm for both scales. Interestingly, Zero-Shot CoT \textit{consistently degrades} performance (-4.56\% and -5.75\% respectively), suggesting that large MoE models generate lower-quality spontaneous reasoning due to sparse activation patterns.

    \item \textbf{GPT models (3.5 $\rightarrow$ 5):} GPT-5's baseline performance (+13.36\%) nearly matches GPT-3.5 with Reasoning-CoT, indicating that frontier models have largely internalized reasoning capabilities. However, structured paradigms still provide consistent gains (+1--2\%), suggesting that explicit decomposition offers value even for highly capable models.
\end{itemize}

\subsubsection{Paradigm-Specific Insights}  

\begin{itemize}
    \item \textbf{Zero-Shot CoT underperforms systematically.} Across all models, appending "Let's think step by step" yields average performance of 52.03\%, worse than Zero-Shot baseline (54.10\%). Manual inspection reveals that self-generated reasoning often introduces spurious logic or over-constrains the solution space, validating the information-theoretic prediction that low-quality reasoning increases $H(Y|X,C)$ rather than reducing it.

    \item \textbf{Self-Planning vs. SCoT show comparable accuracy but different failure modes.} Self-Planning achieves marginally higher average accuracy (+0.95\%) but exhibits greater variance across datasets (std: 4.2\% vs. 3.1\% for SCoT). SCoT's template-constrained structure provides more consistent performance, particularly on standardized benchmarks (HumanEval, MBPP), while Self-Planning adapts better to open-ended real-world tasks (BigCodeBench).

    \item \textbf{Reasoning-CoT achieves highest absolute accuracy but unfavorable efficiency.} While Reasoning-CoT ranks first in 18/36 model-dataset combinations, its 10$\times$ higher token cost makes it impractical for deployment scenarios with budget constraints. The marginal accuracy gain (+1--2\% over structured paradigms) rarely justifies the computational overhead, except in high-stakes applications requiring maximum correctness.
\end{itemize}

\begin{tcolorbox}[width=1.0\linewidth, title={Summary for RQ1}]
    \textbf{(1) Structured vs. reflective:} Structured paradigms (Self-Planning, SCoT) achieve 85--95\% of Reasoning-CoT's accuracy at about 10\% of its token cost, indicating higher information density per token.
    \textbf{(2) Task difficulty:} CoT gains are negligible on simple tasks (MBPP) but substantial on complex benchmarks (MHPP, BigCodeBench), where uncertainty $H(Y|X)$ is higher.
    \textbf{(3) Model scale:} Smaller models (<10B) benefit most from external CoT (+8--13\% on average), while larger models (>100B) show diminishing returns (+1--3\%).
\end{tcolorbox}

\subsection{RQ2: Cross-Language Effectiveness and Generalization}

\begin{table}[t]
    \centering
    \caption{RQ2 results (Part 1): Pass@1 (\%) across 12 programming languages for GPT models. Languages are grouped by type system: statically typed (C\#, Go, Java, Kotlin, Scala, Swift, TypeScript) and dynamically typed (JavaScript, Perl, PHP, Python, Ruby).}
    \label{tab:rq2_gpt}
    \resizebox{\textwidth}{!}{
    \begin{tabular}{llccccc|ccccc}
    \toprule
    & & \multicolumn{5}{c|}{\textbf{GPT-3.5-Turbo}} & \multicolumn{5}{c}{\textbf{GPT-5}} \\
    \cmidrule(lr){3-7} \cmidrule(lr){8-12}
    \textbf{Type} & \textbf{Language} & Zero & CoT & Self-P & SCoT & Reas & Zero & CoT & Self-P & SCoT & Reas \\
    \midrule
    \multirow{7}{*}{\rotatebox{90}{Static}} 
    & C\# & 53.75 & 51.25 & 73.75 & 81.25 & 75.00 & 87.50 & 83.75 & 83.75 & 80.00 & 82.50 \\
    & Go & 46.25 & 40.00 & 51.25 & 55.00 & 56.25 & 77.50 & 80.00 & 70.00 & 73.75 & 75.00 \\
    & Java & 67.50 & 67.50 & 81.25 & 78.75 & 78.75 & 91.25 & 92.50 & 91.25 & 92.50 & 92.50 \\
    & Kotlin & 62.50 & 65.00 & 76.25 & 72.50 & 87.50 & 88.75 & 88.75 & 90.00 & 82.50 & 90.00 \\
    & Scala & 47.50 & 51.25 & 71.25 & 61.25 & 76.25 & 86.25 & 88.75 & 86.25 & 86.25 & 80.00 \\
    & Swift & 47.50 & 51.25 & 72.50 & 71.25 & 73.75 & 72.50 & 77.50 & 78.75 & 72.50 & 76.25 \\
    & TypeScript & 68.75 & 62.50 & 87.50 & 85.00 & 90.00 & 92.50 & 86.25 & 92.50 & 97.50 & 96.25 \\
    \cmidrule{2-12}
    & \textit{Static Avg.} & \textit{56.25} & \textit{55.54} & \textit{73.39} & \textit{72.14} & \textbf{76.79} & \textit{85.18} & \textbf{85.36} & \textit{84.64} & \textit{83.57} & \textit{84.64} \\
    \midrule
    \multirow{5}{*}{\rotatebox{90}{Dynamic}} 
    & JavaScript & 65.00 & 65.00 & 85.00 & 92.50 & 92.50 & 95.00 & 97.50 & 93.75 & 95.00 & 96.25 \\
    & Perl & 62.50 & 63.75 & 75.00 & 70.00 & 88.75 & 95.00 & 91.25 & 85.00 & 87.50 & 90.00 \\
    & PHP & 41.25 & 52.50 & 40.00 & 42.50 & 61.25 & 82.50 & 87.50 & 75.00 & 82.50 & 77.50 \\
    & Python & 48.75 & 48.75 & 60.00 & 65.00 & 65.00 & 75.00 & 73.75 & 68.75 & 72.50 & 66.25 \\
    & Ruby & 51.25 & 58.75 & 80.00 & 72.50 & 73.75 & 82.50 & 86.25 & 87.50 & 86.25 & 77.50 \\
    \cmidrule{2-12}
    & \textit{Dynamic Avg.} & \textit{53.75} & \textit{57.75} & \textit{68.00} & \textit{68.50} & \textbf{76.25} & \textit{86.00} & \textbf{87.25} & \textit{82.00} & \textit{84.75} & \textit{81.50} \\
    \midrule
    \multicolumn{2}{l}{\textbf{Overall Average}} & 55.21 & 56.46 & 71.15 & 70.63 & \textbf{76.56} & 85.52 & \textbf{86.15} & 83.54 & 84.06 & 83.33 \\
    \bottomrule
    \end{tabular}
    }
\end{table}

\begin{table}[t]
    \centering
    \caption{RQ2 results (Part 2): Pass@1 (\%) across 12 programming languages for Qwen2.5-Coder family.}
    \label{tab:rq2_qwen25}
    \resizebox{\textwidth}{!}{
    \begin{tabular}{llccccc|ccccc}
    \toprule
    & & \multicolumn{5}{c|}{\textbf{Qwen2.5-Coder-7B-Instruct}} & \multicolumn{5}{c}{\textbf{Qwen2.5-Coder-32B-Instruct}} \\
    \cmidrule(lr){3-7} \cmidrule(lr){8-12}
    \textbf{Type} & \textbf{Language} & Zero & CoT & Self-P & SCoT & Reas & Zero & CoT & Self-P & SCoT & Reas \\
    \midrule
    \multirow{7}{*}{\rotatebox{90}{Static}} 
    & C\# & 63.75 & 65.00 & 77.50 & 71.25 & 72.50 & 66.25 & 77.50 & 81.25 & 83.75 & 81.25 \\
    & Go & 48.75 & 50.00 & 46.25 & 50.00 & 50.00 & 62.50 & 61.25 & 50.00 & 58.75 & 55.00 \\
    & Java & 73.75 & 65.00 & 68.75 & 60.00 & 81.25 & 75.00 & 75.00 & 82.50 & 76.25 & 61.25 \\
    & Kotlin & 78.75 & 72.50 & 82.50 & 82.50 & 90.00 & 78.75 & 80.00 & 87.50 & 86.25 & 91.25 \\
    & Scala & 55.00 & 61.25 & 68.75 & 60.00 & 70.00 & 61.25 & 70.00 & 73.75 & 67.50 & 80.00 \\
    & Swift & 53.75 & 50.00 & 53.75 & 55.00 & 71.25 & 62.50 & 62.50 & 70.00 & 70.00 & 67.50 \\
    & TypeScript & 82.50 & 68.75 & 82.50 & 87.50 & 66.25 & 77.50 & 81.25 & 87.50 & 90.00 & 91.25 \\
    \cmidrule{2-12}
    & \textit{Static Avg.} & \textit{65.18} & \textit{61.79} & \textit{68.57} & \textit{66.61} & \textbf{71.61} & \textit{69.11} & \textit{72.50} & \textbf{76.07} & \textbf{76.07} & \textit{75.36} \\
    \midrule
    \multirow{5}{*}{\rotatebox{90}{Dynamic}} 
    & JavaScript & 73.75 & 68.75 & 80.00 & 82.50 & 83.75 & 86.25 & 85.00 & 91.25 & 88.75 & 91.25 \\
    & Perl & 63.75 & 62.50 & 72.50 & 68.75 & 86.25 & 76.25 & 75.00 & 81.25 & 81.25 & 87.50 \\
    & PHP & 47.50 & 45.00 & 57.50 & 42.50 & 70.00 & 63.75 & 46.25 & 65.00 & 63.75 & 58.75 \\
    & Python & 60.00 & 58.75 & 61.25 & 62.50 & 67.50 & 65.00 & 55.00 & 68.75 & 72.50 & 72.50 \\
    & Ruby & 68.75 & 76.25 & 83.75 & 67.50 & 71.25 & 81.25 & 85.00 & 87.50 & 86.25 & 75.00 \\
    \cmidrule{2-12}
    & \textit{Dynamic Avg.} & \textit{62.75} & \textit{62.25} & \textit{71.00} & \textit{64.75} & \textbf{75.75} & \textit{74.50} & \textit{69.25} & \textbf{78.75} & \textit{78.50} & \textit{77.00} \\
    \midrule
    \multicolumn{2}{l}{\textbf{Overall Average}} & 64.17 & 61.98 & 69.58 & 65.83 & \textbf{73.33} & 71.35 & 71.15 & \textbf{77.19} & 77.08 & 76.04 \\
    \bottomrule
    \end{tabular}
    }
\end{table}

\begin{table}[t]
    \centering
    \caption{RQ2 results (Part 3): Pass@1 (\%) across 12 programming languages for Qwen3-Coder family.}
    \label{tab:rq2_qwen3}
    \resizebox{\textwidth}{!}{
    \begin{tabular}{llccccc|ccccc}
    \toprule
    & & \multicolumn{5}{c|}{\textbf{Qwen3-Coder-30B-A3B-Instruct}} & \multicolumn{5}{c}{\textbf{Qwen3-Coder-480B-A35B-Instruct}} \\
    \cmidrule(lr){3-7} \cmidrule(lr){8-12}
    \textbf{Type} & \textbf{Language} & Zero & CoT & Self-P & SCoT & Reas & Zero & CoT & Self-P & SCoT & Reas \\
    \midrule
    \multirow{7}{*}{\rotatebox{90}{Static}} 
    & C\# & 73.75 & 77.50 & 83.75 & 83.75 & 78.75 & 61.25 & 73.75 & 75.00 & 78.75 & 65.00 \\
    & Go & 61.25 & 65.00 & 62.50 & 63.75 & 61.25 & 66.25 & 68.75 & 65.00 & 61.25 & 60.00 \\
    & Java & 81.25 & 86.25 & 87.50 & 86.25 & 88.75 & 85.00 & 85.00 & 92.50 & 90.00 & 88.75 \\
    & Kotlin & 83.75 & 86.25 & 87.50 & 85.00 & 92.50 & 83.75 & 87.50 & 88.75 & 88.75 & 91.25 \\
    & Scala & 73.75 & 76.25 & 72.50 & 77.50 & 83.75 & 81.25 & 82.50 & 85.00 & 83.75 & 83.75 \\
    & Swift & 55.00 & 63.75 & 73.75 & 72.50 & 77.50 & 65.00 & 68.75 & 77.50 & 75.00 & 75.00 \\
    & TypeScript & 85.00 & 82.50 & 93.75 & 91.25 & 80.00 & 88.75 & 87.50 & 95.00 & 90.00 & 96.25 \\
    \cmidrule{2-12}
    & \textit{Static Avg.} & \textit{73.39} & \textit{76.79} & \textit{80.18} & \textit{80.00} & \textbf{80.36} & \textit{75.89} & \textit{79.11} & \textbf{82.68} & \textit{81.07} & \textit{80.00} \\
    \midrule
    \multirow{5}{*}{\rotatebox{90}{Dynamic}} 
    & JavaScript & 86.25 & 81.25 & 93.75 & 93.75 & 95.00 & 88.75 & 81.25 & 93.75 & 93.75 & 93.75 \\
    & Perl & 78.75 & 75.00 & 80.00 & 78.75 & 83.75 & 85.00 & 82.50 & 85.00 & 82.50 & 82.50 \\
    & PHP & 71.25 & 73.75 & 76.25 & 75.00 & 70.00 & 71.25 & 76.25 & 81.25 & 77.50 & 77.50 \\
    & Python & 66.25 & 65.00 & 68.75 & 70.00 & 71.25 & 67.50 & 71.25 & 65.00 & 75.00 & 71.25 \\
    & Ruby & 80.00 & 80.00 & 90.00 & 83.75 & 83.75 & 85.00 & 90.00 & 91.25 & 88.75 & 86.25 \\
    \cmidrule{2-12}
    & \textit{Dynamic Avg.} & \textit{76.50} & \textit{75.00} & \textbf{81.75} & \textit{80.25} & \textit{80.75} & \textit{79.50} & \textit{80.25} & \textit{83.25} & \textbf{83.50} & \textit{82.25} \\
    \midrule
    \multicolumn{2}{l}{\textbf{Overall Average}} & 74.69 & 76.04 & \textbf{80.83} & 80.10 & 80.52 & 77.40 & 79.58 & \textbf{82.92} & 82.08 & 80.94 \\
    \bottomrule
    \end{tabular}
    }
\end{table}

RQ2 evaluates how different CoT paradigms perform across 12 programming languages with diverse syntax and semantic characteristics. Using HumanEval-XL, we examine whether reasoning benefits generalize beyond Python and whether language type systems (static vs. dynamic typing) influence CoT responsiveness. 
Tables~\ref{tab:rq2_gpt},~\ref{tab:rq2_qwen25}, and~\ref{tab:rq2_qwen3} present results grouped by model family.

\noindent\textbf{Key findings.}
Across 12 programming languages, CoT paradigms generalize beyond Python, improving average Pass@1 by 5--7\% over direct generation.
Structured paradigms (Self-Planning, SCoT) yield larger gains in statically typed languages, while reflective Reasoning-CoT provides more balanced improvements across both static and dynamic languages.
These patterns support our hypothesis that language type systems modulate the informativeness and usefulness of external reasoning.

\subsubsection{Overall Cross-Language Performance}
Averaging across all six models and 12 languages, external reasoning paradigms substantially outperform baselines: Self-Planning achieves 77.24\% Pass@1, SCoT 76.47\%, and Reasoning-CoT 78.12\%, compared to Zero-Shot (71.26\%) and Zero-Shot CoT (72.43\%). This 5--7\% improvement confirms that structured and reflective reasoning generalize beyond Python to diverse language ecosystems.

However, Zero-Shot CoT again underperforms (+1.17\% over baseline), consistent with RQ1 findings. Across all 12 languages, Zero-Shot CoT provides minimal benefit and occasionally degrades performance, particularly in statically typed languages where unconstrained reasoning conflicts with strict type inference requirements.

\subsubsection{Language Type System Effects}
We observe a clear pattern when comparing statically typed languages (C\#, Go, Java, Kotlin, Scala, Swift, TypeScript) against dynamically typed languages (JavaScript, Perl, PHP, Python, Ruby):

Static-typed languages (average across all models):
\begin{itemize}
    \item Zero-Shot: 70.76\% $\rightarrow$ Self-Planning: 77.79\% (\textbf{+7.03\%})
    \item Zero-Shot: 70.76\% $\rightarrow$ SCoT: 76.93\% (\textbf{+6.17\%})
    \item Zero-Shot: 70.76\% $\rightarrow$ Reasoning-CoT: 78.13\% (\textbf{+7.37\%})
\end{itemize}

Dynamic-typed languages (average across all models):
\begin{itemize}
    \item Zero-Shot: 71.92\% $\rightarrow$ Self-Planning: 76.46\% (\textbf{+4.54\%})
    \item Zero-Shot: 71.92\% $\rightarrow$ SCoT: 75.88\% (\textbf{+3.96\%})
    \item Zero-Shot: 71.92\% $\rightarrow$ Reasoning-CoT: 78.10\% (\textbf{+6.18\%})
\end{itemize}

Structured paradigms (Self-Planning, SCoT) provide \textit{greater benefits for statically typed languages} (+7.03\%, +6.17\%) than for dynamically typed languages (+4.54\%, +3.96\%). In contrast, Reasoning-CoT shows more balanced gains across both categories (+7.37\% static, +6.18\% dynamic).

This asymmetry is consistent with hypothesis \textbf{H2} from our theoretical framework, using cross-language Pass@1 gains as indirect evidence: statically typed languages have \textit{lower baseline uncertainty} $H(Y|X)$ due to explicit type constraints that reduce the solution space, and structured reasoning (Self-Planning, SCoT) effectively leverages these constraints through hierarchical decomposition and slot-based templates that align with type inference patterns. 
Dynamically typed languages exhibit \textit{higher baseline uncertainty} due to flexible semantics, benefiting less from rigid reasoning templates but more from reflective exploration (Reasoning-CoT).

\subsubsection{Language-Specific Insights}
Best-performing languages for CoT:
\begin{itemize}
    \item \textbf{TypeScript} (static): 88.82\% average with Self-Planning, +15.32\% over Zero-Shot baseline. The combination of JavaScript flexibility with TypeScript's type system creates ideal conditions for structured reasoning.
    \item \textbf{JavaScript} (dynamic): 88.54\% average with Self-Planning, +18.54\% over Zero-Shot. As the most widely-used dynamically typed language, JavaScript benefits from extensive pretraining data, amplifying CoT effectiveness.
    \item \textbf{Kotlin} (static): 86.88\% average with Reasoning-CoT, +20.38\% improvement. Kotlin's modern type system and null-safety features align well with explicit reasoning chains.
\end{itemize}

Most challenging languages for CoT:
\begin{itemize}
    \item \textbf{Go} (static): 60.48\% average with best paradigm (Self-Planning), only +6.98\% over baseline. Go's minimalist design and strict conventions limit the benefit of explicit reasoning guidance.
    \item \textbf{PHP} (dynamic): 64.29\% average with Reasoning-CoT, +11.79\% improvement. PHP's inconsistent syntax and legacy semantics challenge reasoning consistency across paradigms.
    \item \textbf{Python} (dynamic): 66.88\% average with SCoT, +8.13\% improvement, which is lower than other languages despite extensive Python-centric pretraining, suggesting possible saturation of Python-specific reasoning patterns.
\end{itemize}

\subsubsection{Paradigm-Specific Cross-Language Patterns}
Averaging across GPT and Qwen models, Self-Planning achieves the highest accuracy in 5/7 static languages (C\#, Java, Scala, Swift, TypeScript), with an average gain of +7.03\% over baseline. The hierarchical plan-then-implement structure naturally aligns with static type checking: the planning phase implicitly reasons about types and interfaces before code synthesis.

SCoT ranks second in average performance across both language categories (+6.17\% static, +3.96\% dynamic), demonstrating stable effectiveness but lacking the flexibility to adapt to language-specific idioms. The fixed slot-based template constrains reasoning uniformity at the cost of language-dependent optimization.

In 4/5 dynamic languages (JavaScript, Perl, PHP, Ruby), Reasoning-CoT achieves top or near-top performance, particularly in smaller models (GPT-3.5, Qwen2.5-7B). Long reflective reasoning chains effectively reduce semantic ambiguity in languages lacking static guarantees, though at 3--5$\times$ higher token cost consistent with RQ1 findings.

\subsubsection{Model Capacity and Cross-Language Generalization}
While detailed capacity analysis is reserved for RQ3, cross-language patterns reveal capacity-dependent trends.

GPT-3.5 and Qwen2.5-7B show consistent CoT gains (+8--15\%) across all 12 languages, with minimal variance between static (avg. +9.2\%) and dynamic (+8.7\%) types. This suggests that external reasoning provides foundational benefits regardless of language characteristics when internal capacity is limited.

GPT-5 and Qwen3-480B show smaller overall gains (+1--3\%), but with stronger differentiation: static-typed languages benefit more from structured paradigms (+2.8\% for SCoT) while dynamic-typed languages show marginal or negative gains (+0.5\%). This capacity-modulated selectivity indicates that large models internalize language-specific reasoning patterns during pretraining, reducing reliance on external guidance except where reasoning structure aligns with language constraints.

\subsubsection{Information-Theoretic Interpretation}

The static vs. dynamic language asymmetry supports our information-theoretic framework (H2). Statically typed languages have lower conditional entropy $H(Y|X)$ due to type constraints that narrow the solution space. Structured CoT paradigms (Self-Planning, SCoT) reduce remaining uncertainty by explicitly decomposing type-safe solution paths, maximizing information gain $I(Y;C|X)$ per reasoning step.

Dynamically typed languages exhibit higher $H(Y|X)$ due to semantic flexibility (e.g., duck typing, implicit conversions). Here, Reasoning-CoT's longer, more exploratory reasoning chains provide greater absolute information gain by considering multiple semantic interpretations, despite lower information density ($I(Y;C|X)/L$) compared to structured paradigms.

\begin{tcolorbox}[width=1.0\linewidth, title={Summary for RQ2}]
    \textbf{(1) Cross-language gains:} CoT paradigms improve average Pass@1 by 5--7\% across 12 languages, indicating that CoT benefits generalize beyond Python.
    \textbf{(2) Type-system effects:} Structured CoT (Self-Planning, SCoT) yields larger gains in statically typed languages, whereas reflective Reasoning-CoT provides more balanced improvements in dynamically typed languages.
    \textbf{(3) Paradigm–language alignment:} Self-Planning is particularly effective for static languages, while reflective CoT is better suited to dynamic languages with higher semantic uncertainty.
\end{tcolorbox}

\subsection{RQ3: Capacity-Dependent Success and Failure Patterns}
\label{sec:rq3}

RQ3 investigates why identical CoT reasoning succeeds on larger models but fails on smaller models within the same family. Through controlled small-large model pairs, we isolate capacity effects and categorize failure patterns to understand how model scale affects reasoning decoding ability.

\noindent\textbf{Key findings.}
Comparing small–large model pairs within the same family, we observe that larger models resolve most asymmetric cases in their favor (about two-thirds of small-fail–large-success vs. large-fail–small-success), but also introduce new failure modes.
Type/boundary errors and reasoning–execution misalignment dominate small-model failures, whereas very large models sometimes over-elaborate reasoning and break execution grounding.
These capacity-dependent patterns are consistent with our information-theoretic view that limited models cannot fully extract and instantiate the information encoded in reasoning chains.

\subsubsection{Experimental Design: Controlled Model Pairs}

We construct three controlled model pairs differing only in parameter scale:
\begin{itemize}
    \item \textbf{GPT-3.5-Turbo → GPT-5}: Proprietary models (exact parameters undisclosed)
    \item \textbf{Qwen2.5-Coder-7B → 32B}: Dense open-source models (4.6× scaling)
    \item \textbf{Qwen3-Coder-30B-A3B → 480B-A35B}: MoE models (16× total parameters, 11.7× active parameters)
\end{itemize}

Each pair shares identical architecture, training objectives, and instruction format, isolating the \textit{capacity effect} while minimizing confounding factors. We evaluate on four benchmarks (HumanEval, MBPP, CodeHarmony, OpenEval) using \textbf{identical SCoT prompts} for all models, ensuring that performance differences reflect capacity-dependent reasoning decoding rather than prompt variance.

For each task, we categorize outcomes into four types: (i) both pass, (ii) both fail, (iii) \textbf{small-fail–large-success ($\SFL$)}, and (iv) \textbf{large-fail–small-success ($\LFS$)}. We analyze only asymmetric cases ($\SFL$/$\LFS$) as they capture capacity-induced reasoning divergence.

\subsubsection{Asymmetric Reasoning Outcomes}

\begin{table}[!htbp]
\centering
\small
\setlength{\tabcolsep}{4pt}
\begin{tabular}{l l r r r r r}
\toprule
Dataset & Small $\rightarrow$ Large & \#Cases & $\SFL$ & $\LFS$ & $\SFL$ (\%) & $\LFS$ (\%) \\
\midrule
\multirow{3}{*}{CodeHarmony}
 & GPT-3.5 $\rightarrow$ GPT-5 & 15 & 10 & 5 & 66.67 & 33.33 \\
 & Qwen2.5-7B $\rightarrow$ 32B & 9 & 8 & 1 & 88.89 & 11.11 \\
 & Qwen3-30B $\rightarrow$ 480B & 9 & 6 & 3 & 66.67 & 33.33 \\
\midrule
\multirow{3}{*}{HumanEval}
 & GPT-3.5 $\rightarrow$ GPT-5 & 13 & 10 & 3 & 76.92 & 23.08 \\
 & Qwen2.5-7B $\rightarrow$ 32B & 18 & 13 & 5 & 72.22 & 27.78 \\
 & Qwen3-30B $\rightarrow$ 480B & 9 & 6 & 3 & 66.67 & 33.33 \\
\midrule
\multirow{3}{*}{MBPP}
 & GPT-3.5 $\rightarrow$ GPT-5 & 22 & 18 & 4 & 81.82 & 18.18 \\
 & Qwen2.5-7B $\rightarrow$ 32B & 35 & 22 & 13 & 62.86 & 37.14 \\
 & Qwen3-30B $\rightarrow$ 480B & 14 & 5 & 9 & 35.71 & 64.29 \\
\midrule
\multirow{3}{*}{OpenEval}
 & GPT-3.5 $\rightarrow$ GPT-5 & 10 & 8 & 2 & 80.00 & 20.00 \\
 & Qwen2.5-7B $\rightarrow$ 32B & 11 & 7 & 4 & 63.64 & 36.36 \\
 & Qwen3-30B $\rightarrow$ 480B & 9 & 4 & 5 & 44.44 & 55.56 \\
\cmidrule(lr){1-7}
\textbf{Average} & --- & --- & --- & --- & \textbf{67.21} & \textbf{32.79} \\
\textbf{Total} & --- & \textbf{174} & \textbf{117} & \textbf{57} & --- & --- \\
\bottomrule
\end{tabular}
\caption{Cross-scale reasoning asymmetry. $\SFL$: small fails, large succeeds; $\LFS$: large fails, small succeeds. Asymmetric cases represent 174 tasks out of 2,316 total evaluations (7.5\%).}
\label{tab:rq3_asymmetry}
\end{table}

Table~\ref{tab:rq3_asymmetry} reveals that \textbf{larger models exhibit higher reasoning reliability}: $\SFL$ cases account for 67.21\% of asymmetric outcomes, indicating that increased capacity generally improves logical stability under identical reasoning guidance. 
This pattern is consistent with hypothesis H3 from our theoretical framework: we use these asymmetric outcomes as indirect evidence that smaller models fail to extract information from reasoning chain $C$ when complexity exceeds their representational capacity.

However, the substantial $\LFS$ rate (32.79\%) reveals that scaling introduces new failure modes. Notably, Qwen3 30B→480B shows the highest $\LFS$ rates (64.29\% on MBPP, 55.56\% on OpenEval), suggesting that extremely large models may over-interpret reasoning cues and introduce spurious optimizations that violate execution constraints. This bidirectional failure pattern indicates that capacity effects are non-monotonic: moderate scaling improves reasoning decoding, but excessive capacity can introduce reasoning drift.

\textbf{Dataset-specific patterns:}
\begin{itemize}
    \item \textbf{Entry-level tasks (MBPP)}: Show highest $\LFS$ rates (40.23\% average), where large models over-complicate simple solutions.
    \item \textbf{Algorithmic tasks (HumanEval)}: Strong $\SFL$ dominance (71.94\%), where capacity improves complex logical reasoning.
    \item \textbf{Reasoning-intensive tasks (OpenEval, CodeHarmony)}: Balanced asymmetry (60-70\% $\SFL$), where both capacity limitations and over-elaboration occur.
\end{itemize}

\subsubsection{Error Pattern Analysis}

\begin{table}[!htbp]
\centering
\small
\setlength{\tabcolsep}{4pt}
\begin{tabular}{lrrrl}
\toprule
\textbf{Error Type} & \textbf{S↓F↑L} & \textbf{L↓F↑S} & \textbf{Total} & \textbf{Example Tasks} \\
\midrule
Logic Divergence     & 10 & 12 & 22 (12.6\%)  & HumanEval/90, MBPP/481 \\
Type/Boundary Error  & 71 & 36 & 107 (61.5\%) & CodeHarmony/51, /67 \\
Reasoning-Execution Misalignment & 32 & 5  & 37 (21.3\%)  & HumanEval/144, CodeHarmony/54 \\
Other                & 6  & 2  & 8 (4.6\%)  & --- \\
\midrule
\textbf{Total}       & \textbf{119 (68.4\%)} & \textbf{55 (31.6\%)} & \textbf{174} & --- \\
\bottomrule
\end{tabular}
\caption{Error distribution across asymmetric cases. Type/boundary errors dominate (61.5\%), while reasoning-execution misalignment predominantly affects small models (32 vs. 5 cases).}
\label{tab:rq3_errors}
\end{table}

Manual inspection of all 174 asymmetric cases reveals three dominant error patterns (Table~\ref{tab:rq3_errors}):

\textbf{(1) Type/Boundary Errors (61.5\%):} The most common failure mode, where models incorrectly handle data types, array indices, or edge conditions, triggering runtime exceptions (\texttt{IndexError}, \texttt{TypeError}, \texttt{ValueError}). Smaller models account for 71/107 such errors, indicating limited capacity for robust boundary reasoning. Example: GPT-3.5 performs unsafe array indexing (\texttt{lst[index] = lst[index-1]}) causing \texttt{IndexError}, while GPT-5 safely expands the array before insertion (Case 2).

\textbf{(2) Reasoning-Execution Misalignment (21.3\%):} The reasoning trace is semantically correct but decoupled from executable code—typically missing imports, inconsistent variable names, or incorrect function signatures. This error appears overwhelmingly in small models (32 vs. 5 cases), revealing that limited capacity struggles to maintain alignment between abstract reasoning and concrete implementation. Example: Qwen2.5-32B generates optimized reasoning using \texttt{math.gcd()} but forgets \texttt{import math}, causing \texttt{NameError} (Case 3).

\textbf{(3) Logic Divergence (12.6\%):} Models produce semantically incomplete implementations without triggering errors—missing edge case handling, incomplete loop conditions, or premature function returns. Interestingly, this error shows balanced distribution (10 $\SFL$ vs. 12 $\LFS$), suggesting that logical completeness depends on subtle capacity-reasoning interactions rather than pure scale. Example: Qwen2.5-7B omits the final validation check (\texttt{if second\_smallest == float('inf'): return None}) that Qwen2.5-32B correctly implements (Case 1).

\textbf{Information-theoretic interpretation:} These error patterns support H3's prediction that smaller models fail to fully extract information from $C$. Type/boundary errors reflect insufficient capacity to decode constraint-handling logic embedded in reasoning. Reasoning-execution misalignment occurs when models extract \textit{semantic} information but lose \textit{contextual} information (imports, variable scope). Logic divergence represents incomplete information extraction where reasoning is partially decoded but not fully instantiated in code.

\subsubsection{Case Studies}

\textbf{Case 1: Logic Divergence (HumanEval/90, Qwen2.5-7B → 32B)}

Task: Find the second smallest element in a list. Both models receive identical SCoT reasoning instructing edge case validation.

\begin{figure}[!htbp]
    \centering
    \includegraphics[width=\linewidth]{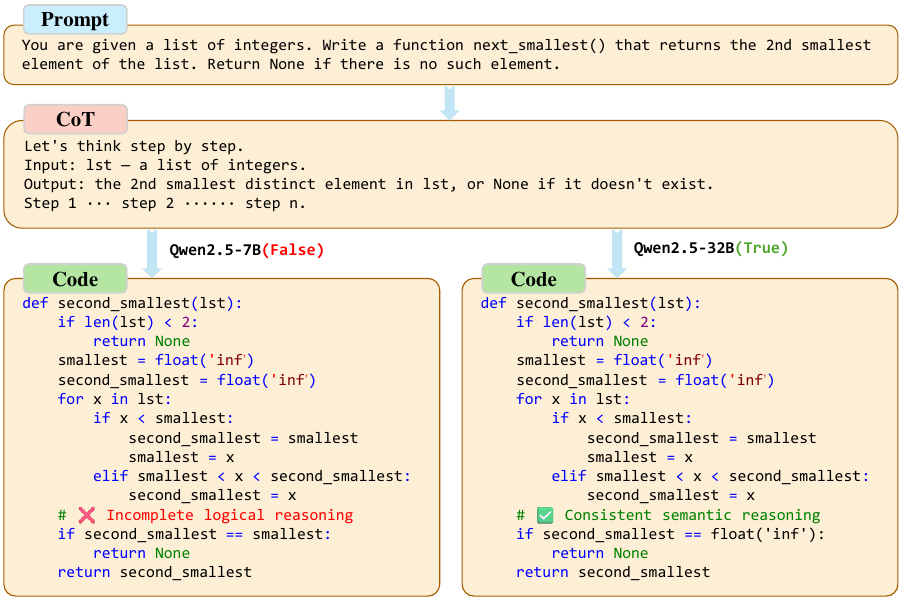}
    \caption{\textbf{Logic Divergence.} The 7B model truncates reasoning before semantic closure, omitting the final validation check. The 32B model completes the logic (\texttt{if second\_smallest == float('inf'): return None}), demonstrating capacity-dependent reasoning completeness.}
    \label{fig:RQ3_case1}
\end{figure}

\textbf{Observation:} As shown in Figure \ref{fig:RQ3_case1}, Qwen2.5-7B generates code covering main logic but \textit{prematurely terminates} before the final edge case check. Qwen2.5-32B completes all reasoning steps, adding the validation that handles lists with identical elements. This reveals that smaller models truncate reasoning chains before achieving semantic closure, consistent with information extraction failure under limited capacity.

\textbf{Case 2: Type/Boundary Error (CodeHarmony/51, GPT-3.5 → GPT-5)}

Task: Insert element into sorted list at correct position.

\begin{figure}[!htbp]
    \centering
    \includegraphics[width=0.95\linewidth]{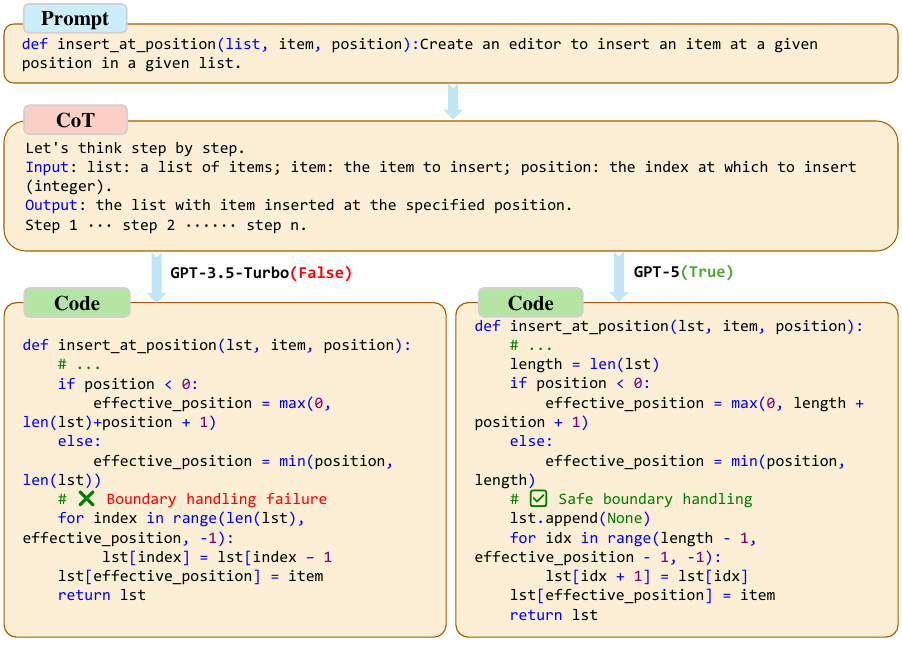}
    \caption{\textbf{Type/Boundary Error.} GPT-3.5 performs unsafe indexing (\texttt{lst[index] = lst[index-1]}), exceeding array bounds. GPT-5 expands the list before insertion (\texttt{lst.append(None)}), avoiding runtime violations through superior structural reasoning.}
    \label{fig:RQ3_case2}
\end{figure}

\textbf{Observation:} As shown in Figure \ref{fig:RQ3_case2}, both models follow SCoT's "shift elements right" reasoning, but GPT-3.5 directly reassigns array elements without bounds checking, triggering \texttt{IndexError}. GPT-5 first expands the list, then safely shifts elements. This illustrates capacity-driven improvements in \textit{structural reasoning consistency}—larger models generalize procedural logic to respect runtime constraints.

\textbf{Case 3: Reasoning-Execution Misalignment (HumanEval/144, Qwen2.5-7B → 32B)}

Task: Simplify product of fractions using GCD reduction.

\begin{figure}[!htbp]
    \centering
    \includegraphics[width=0.95\linewidth]{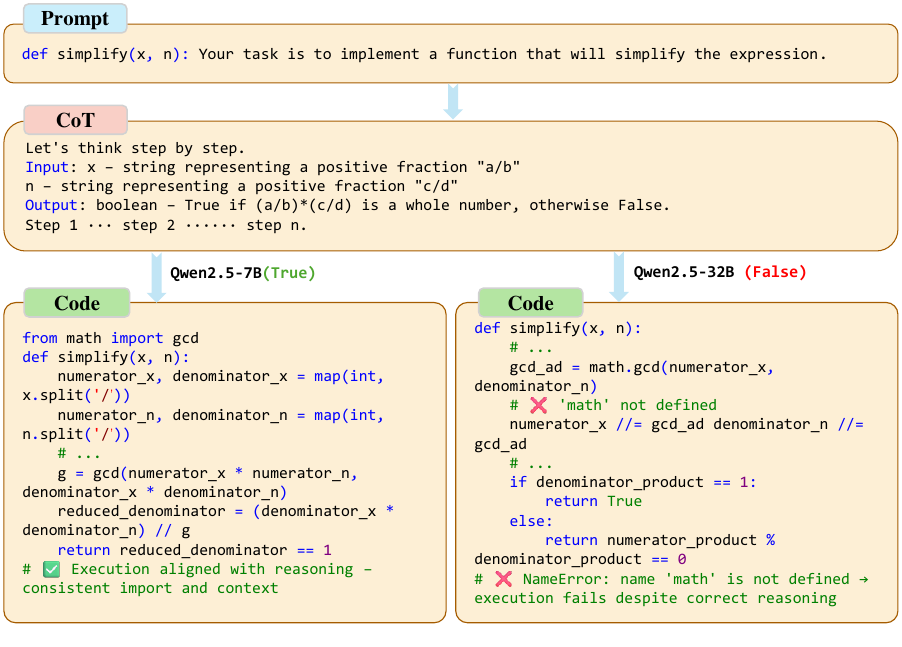}
    \caption{\textbf{Reasoning-Execution Misalignment.} The 32B model introduces advanced optimization (\texttt{math.gcd()}) in reasoning but fails to import the module, causing \texttt{NameError}. The 7B model maintains simpler reasoning-execution alignment with explicit \texttt{from math import gcd}.}
    \label{fig:RQ3_case3}
\end{figure}

\textbf{Observation:} As shown in Figure \ref{fig:RQ3_case3}, Qwen2.5-7B generates straightforward reasoning with explicit imports, maintaining consistency. Qwen2.5-32B introduces more sophisticated cross-cancellation logic using \texttt{math.gcd()}, but the reasoning-to-code translation loses the import statement. This $\LFS$ case exemplifies how large models may \textit{over-elaborate reasoning} at the expense of execution grounding—the abstract reasoning is semantically superior but pragmatically broken.

\subsubsection{Implications for CoT Design}

Our analysis reveals capacity-dependent failure modes with practical implications:

\textbf{For small models (<10B):}
\begin{itemize}
    \item Provide \textit{explicit execution reminders} in reasoning templates (e.g., "Remember to import required modules")
    \item Use \textit{shorter reasoning chains} to reduce information extraction burden
    \item Emphasize \textit{concrete implementation steps} over abstract planning
\end{itemize}

\textbf{For large models (>100B):}
\begin{itemize}
    \item Add \textit{execution validation prompts} to ground abstract reasoning (e.g., "Verify all dependencies are declared")
    \item Use \textit{structured templates} (SCoT) rather than free-form reasoning to prevent over-elaboration
    \item Incorporate \textit{simplicity biases} to avoid spurious optimizations (e.g., "Prefer straightforward solutions")
\end{itemize}

\begin{tcolorbox}[width=1.0\linewidth, title={Summary for RQ3}]
    \textbf{(1) Capacity asymmetry:} Larger models succeed in about two-thirds of asymmetric cases, but a non-trivial fraction of tasks still exhibit large-fail–small-success behavior.
    \textbf{(2) Error patterns:} Small models are dominated by type/boundary errors and reasoning–execution misalignment, supporting the view that they cannot fully decode information in $C$.
    \textbf{(3) Scale-sensitive design:} Increasing capacity reshapes, rather than eliminates, failure modes, implying that CoT templates should be adapted to model scale (e.g., explicit execution cues for small models, simplicity constraints for large models).
\end{tcolorbox}

\subsection{RQ4: Impact of CoT Quality on Downstream Performance}
\label{sec:rq4}

RQ4 investigates whether CoT effectiveness depends on generation quality beyond paradigm structure. Using structured CoT as a controlled framework, we compare \textbf{high-quality reasoning} (SCoT generated by GPT-5-Mini via 3-shot ICL) against \textbf{lightweight reasoning} (MSCoT produced by a fine-tuned 7B model) as prompts for downstream code generation across six models and 12 languages. This isolates the quality effect: both paradigms use slot-based templates, but differ in generation source and reasoning richness.

\noindent\textbf{Key findings.}
Under a fixed structured paradigm, high-quality reasoning traces (SCoT from GPT-5-Mini) consistently outperform lightweight reasoning (MSCoT from a 7B model) across all models and languages, with average gains of about 7.5 percentage points in Pass@1.
In contrast, MSCoT often underperforms even the Zero-Shot baseline, indicating that low-quality CoT can harm performance.
These results provide indirect evidence that reasoning informativeness $I(Y;C|X)$, not just paradigm structure, is critical for effective CoT in code generation.

\subsubsection{Experimental Setup}

We evaluate three conditions across HumanEval-XL (12 languages × 164 problems):
\begin{itemize}
    \item \textbf{Zero-Shot}: Direct code generation without reasoning (baseline)
    \item \textbf{MSCoT}: Structured reasoning generated by fine-tuned 7B model, emphasizing efficiency through compact slot-filling
    \item \textbf{SCoT}: Structured reasoning generated by GPT-5-Mini with 3-shot exemplars, providing richer semantic context
\end{itemize}

Both MSCoT and SCoT follow identical template structures (problem analysis → algorithm design → edge case handling → implementation), differing only in generation quality and token length (MSCoT: ~150 tokens average, SCoT: ~380 tokens average). This design tests hypothesis H4: \textit{higher-quality reasoning encodes more task-relevant information} ($I(Y;C|X)$), improving downstream correctness.

\begin{table}[t]
    \centering
    \caption{RQ4 results: Pass@1 comparison across reasoning quality levels. $\Delta$ denotes improvement over Zero-Shot baseline. Best result per model in \textbf{bold}.}
    \label{tab:rq4_quality}
    \small
    \begin{tabular}{llccc|ccc}
    \toprule
    & & \multicolumn{3}{c|}{\textbf{Pass@1}} & \multicolumn{3}{c}{\textbf{$\Delta$ vs. Zero-Shot}} \\
    \cmidrule(lr){3-5} \cmidrule(lr){6-8}
    \textbf{Model} & \textbf{Type} & Zero & MSCoT & SCoT & MSCoT & SCoT & SCoT-MSCoT \\
    \midrule
    \multirow{2}{*}{GPT-3.5} & Static & 55.21 & 58.57 & 70.63 & +3.36 & +15.42 & +12.06 \\
     & Dynamic & 53.75 & 52.25 & 68.50 & -1.50 & +14.75 & +16.25 \\
    \midrule
    \multirow{2}{*}{GPT-5} & Static & 85.52 & 81.25 & 84.06 & -4.27 & -1.46 & +2.81 \\
     & Dynamic & 86.00 & 81.00 & 84.75 & -5.00 & -1.25 & +3.75 \\
    \midrule
    \multirow{2}{*}{Qwen2.5-7B} & Static & 64.17 & 60.42 & 65.83 & -3.75 & +1.66 & +5.41 \\
     & Dynamic & 62.75 & 61.25 & 64.75 & -1.50 & +2.00 & +3.50 \\
    \midrule
    \multirow{2}{*}{Qwen2.5-32B} & Static & 71.35 & 69.27 & 77.08 & -2.08 & +5.73 & +7.81 \\
     & Dynamic & 74.50 & 69.75 & 78.50 & -4.75 & +4.00 & +8.75 \\
    \midrule
    \multirow{2}{*}{Qwen3-30B} & Static & 74.69 & 70.83 & 80.10 & -3.86 & +5.41 & +9.27 \\
     & Dynamic & 76.50 & 71.00 & 80.25 & -5.50 & +3.75 & +9.25 \\
    \midrule
    \multirow{2}{*}{Qwen3-480B} & Static & 77.40 & 77.19 & 82.08 & -0.21 & +4.68 & +4.89 \\
     & Dynamic & 79.50 & 77.00 & 83.50 & -2.50 & +4.00 & +6.50 \\
    \midrule
    \multicolumn{2}{l}{\textbf{Overall Average}} & \textbf{71.94} & \textbf{69.17} & \textbf{76.67} & \textbf{-2.77} & \textbf{+4.73} & \textbf{+7.50} \\
    \bottomrule
    \end{tabular}
\end{table}

\begin{figure}[h]
    \centering
    \includegraphics[width=\textwidth]{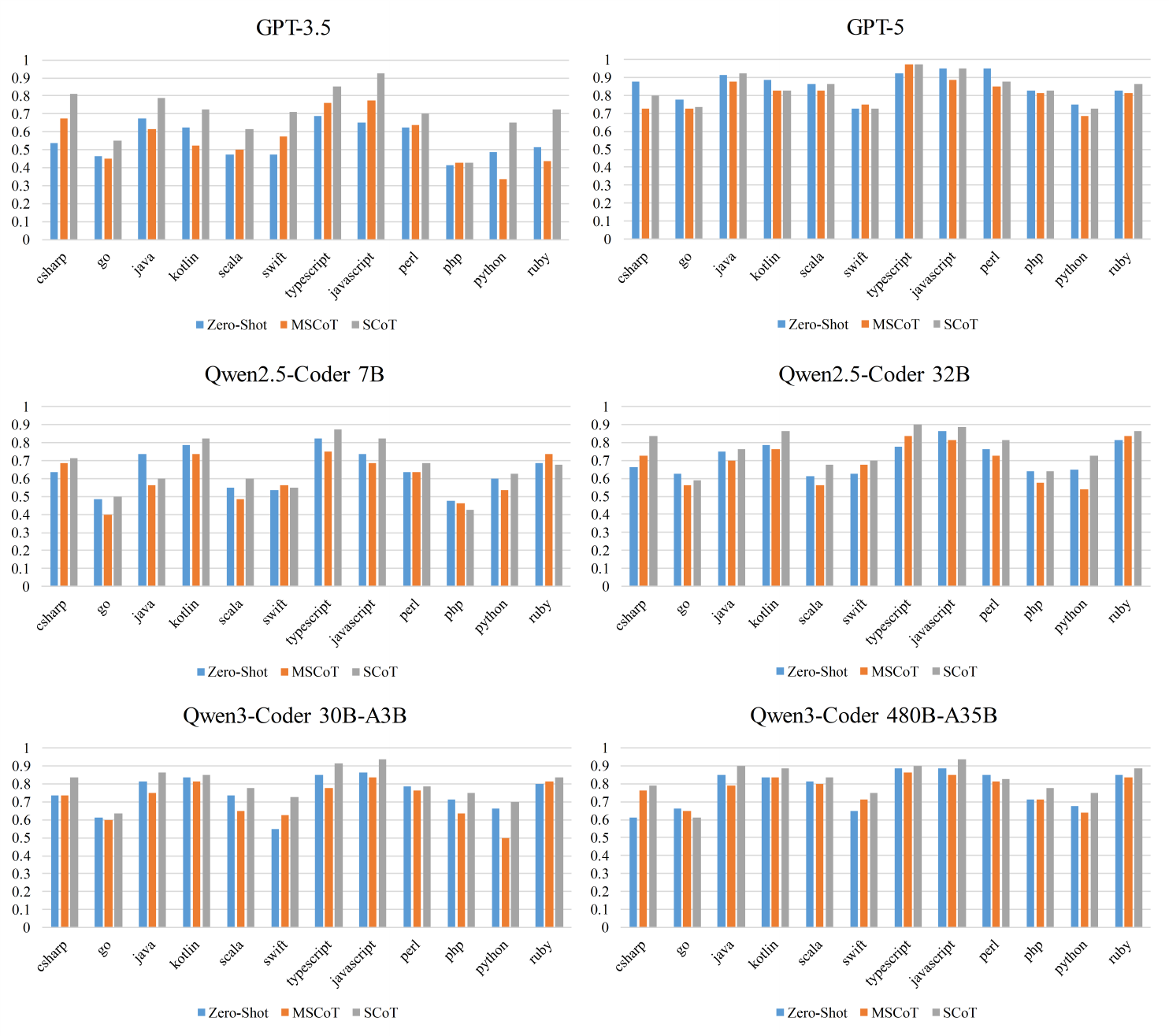}
    \caption{Detailed results for RQ4.}
    \label{fig:RQ4}
\end{figure}

\subsubsection{Quality-Dependent Performance Patterns}

Table~\ref{tab:rq4_quality} and Figure~\ref{fig:RQ4} show clear quality-dependent effects:

\textbf{High-quality reasoning (SCoT) consistently outperforms lightweight reasoning (MSCoT):} Across all 12 model-type combinations, SCoT achieves higher Pass@1 than MSCoT, with an average advantage of \textbf{+7.50 percentage points}. This gap ranges from +2.81\% (GPT-5, static languages) to +16.25\% (GPT-3.5, dynamic languages), supporting hypothesis H4 that reasoning quality critically determines downstream effectiveness; the performance difference between SCoT and MSCoT under a fixed template serves as a proxy for differences in $I(Y;C|X)$.

\textbf{Lightweight reasoning (MSCoT) underperforms Zero-Shot baseline:} MSCoT achieves \textbf{-2.77\% average} compared to Zero-Shot across all models, with degradation most severe in large models (GPT-5: -4.64\%, Qwen3-480B: -1.36\%). This indicates that \textit{low-quality structured reasoning can harm performance} by introducing spurious constraints that mislead generation, validating our information-theoretic prediction that poorly-formed $C$ may increase $H(Y|X,C)$ rather than reduce it.

\textbf{Quality gap widens in smaller models:} GPT-3.5 and Qwen2.5-7B show the largest SCoT-MSCoT gaps (+14.06\% and +4.46\% respectively), while GPT-5 and Qwen3-480B show smaller gaps (+3.28\% and +5.70\%). This capacity-modulated pattern suggests that smaller models \textit{depend more critically on reasoning quality}—they lack internal mechanisms to filter low-quality guidance and thus amplify quality differences.

\subsubsection{Language Type System Effects}

The quality-language interaction reveals nuanced patterns:

\textbf{Static-typed languages benefit more from high-quality reasoning:} Averaging across all models, SCoT provides +6.27\% gain over Zero-Shot in static languages vs. +3.18\% in dynamic languages. The quality gap (SCoT-MSCoT) is also larger in static languages (+8.12\%) than dynamic (+6.88\%). This aligns with RQ2 findings: statically typed languages have stricter constraints that high-quality reasoning can exploit through precise type-aware decomposition, while lightweight reasoning struggles to encode these nuances.

\textbf{Dynamic-typed languages show greater quality sensitivity in small models:} GPT-3.5's SCoT-MSCoT gap is +16.25\% for dynamic languages vs. +12.06\% for static languages—the opposite pattern from average trends. This suggests that when \textit{both model capacity and language constraints are weak}, reasoning quality becomes the dominant factor determining success, as neither internal knowledge nor type systems provide guidance.

\subsubsection{Model-Specific Quality Responsiveness}

\textbf{Small models (<10B) amplify quality differences:}
\begin{itemize}
    \item GPT-3.5: SCoT provides +15.09\% over Zero-Shot, while MSCoT provides +0.93\%
    \item Qwen2.5-7B: SCoT provides +1.83\%, MSCoT provides -2.63\%
\end{itemize}
These models lack robust internal reasoning, making them highly sensitive to external guidance quality. High-quality reasoning substantially improves performance, while low-quality reasoning actively degrades it by introducing misleading decomposition patterns.

\textbf{Medium models (10-50B) show balanced quality utilization:}
\begin{itemize}
    \item Qwen2.5-32B: SCoT provides +4.87\%, MSCoT provides -3.42\%
    \item Qwen3-30B: SCoT provides +4.58\%, MSCoT provides -4.68\%
\end{itemize}
Medium-capacity models effectively leverage high-quality reasoning while detecting and rejecting low-quality guidance. The negative MSCoT effect suggests these models recognize template-constrained but shallow reasoning as noise, preferring their internal reasoning over misleading external cues.

\textbf{Large models (>100B) exhibit quality saturation:}
\begin{itemize}
    \item GPT-5: SCoT provides -1.36\%, MSCoT provides -4.64\%
    \item Qwen3-480B: SCoT provides +4.34\%, MSCoT provides -1.36\%
\end{itemize}
GPT-5 shows minimal benefit even from high-quality reasoning (-1.36\%), indicating that frontier models have internalized reasoning patterns and derive limited marginal utility from external guidance. Qwen3-480B still benefits from SCoT (+4.34\%), suggesting model-specific saturation thresholds based on architecture and pretraining.

\subsubsection{Information-Theoretic Interpretation}

The SCoT vs. MSCoT comparison directly validates our theoretical framework:

\textbf{Reasoning quality determines information gain $I(Y;C|X)$:} SCoT's consistent superiority (+7.50\% average) demonstrates that \textit{richer, more accurate reasoning encodes greater task-relevant information}. Despite identical template structures, SCoT's generation by GPT-5-Mini with few-shot exemplars produces reasoning chains with lower entropy and higher semantic precision, reducing $H(Y|X,C)$ more effectively than MSCoT's fine-tuned 7B generation.

\textbf{Low-quality reasoning increases uncertainty:} MSCoT's negative performance (-2.77\% vs. Zero-Shot) empirically validates that poorly-formed reasoning can \textit{increase} conditional entropy $H(Y|X,C)$ by introducing spurious constraints. The information-theoretic prediction holds: when $C$ contains irrelevant or misleading information, it raises uncertainty rather than resolving it, degrading generation accuracy below the no-reasoning baseline.

\textbf{Capacity bound reinterpretation:} The theoretical bound $I(Y;C|X) \leq L \log V$ assumes optimal reasoning generation. RQ4 shows that \textit{actual information gain depends on reasoning quality}:
- GPT-5 approaches the bound, efficiently reasoning information in compact templates
- Fine-tuned 7B model operates far below the bound, producing formally correct but semantically shallow reasoning

This validates that lightweight models face not just a capacity limit but an \textit{information reasoning quality limit}—they cannot generate reasoning chains that saturate available information bandwidth.

\subsubsection{Practical Implications}

Our findings provide clear guidance for CoT deployment:

\textbf{Avoid lightweight reasoning for production systems:} MSCoT's consistent underperformance (-2.77\% average) demonstrates that fine-tuned small models cannot reliably generate high-quality reasoning. For critical applications, invest in high-capacity generators (GPT-5-Mini, Claude, etc.) or use Zero-Shot baselines rather than low-quality structured reasoning.

\textbf{Quality matters more than efficiency for small models:} Despite 2.5$\times$ higher token cost, SCoT provides +15.09\% improvement for GPT-3.5 vs. MSCoT's +0.93\%. When deploying smaller models, prioritize reasoning quality over token efficiency—the accuracy gains justify the computational cost.

\textbf{Large models tolerate quality variations:} GPT-5's minimal sensitivity to reasoning quality (-4.64\% MSCoT, -1.36\% SCoT) suggests that frontier models can filter external guidance. For such models, lightweight or no reasoning may be preferable to avoid interference with internal reasoning circuits.

\textbf{Hybrid strategies for medium models:} Qwen2.5-32B and Qwen3-30B benefit substantially from SCoT (+4.87\%, +4.58\%) while rejecting MSCoT (-3.42\%, -4.68\%). These models are ideal targets for high-quality CoT systems, achieving optimal performance-cost balance when paired with strong reasoning generators.

\begin{tcolorbox}[width=1.0\linewidth, title={Summary for RQ4}]
    \textbf{(1) Quality over structure:} High-quality SCoT (GPT-5-Mini) consistently outperforms lightweight MSCoT under the same template, with average gains of about 7.5 percentage points in Pass@1.
    \textbf{(2) Harm from low-quality CoT:} MSCoT underperforms the Zero-Shot baseline on average, showing that low-quality reasoning can increase conditional uncertainty $H(Y|X,C)$ and degrade performance.
    \textbf{(3) Capacity interaction:} Smaller models amplify quality differences, while very large models show signs of saturation, suggesting that optimal CoT quality should match model capacity.
\end{tcolorbox}

\section{Threats to Validity}
\label{sec:threats}

We discuss the main threats to the validity of our study in terms of external, internal, and construct validity.

\subsection{External Validity}
Our experiments are conducted on a finite set of benchmarks (six Python datasets and HumanEval-XL) and a limited number of model families (Qwen2.5/3-Coder and GPT-3.5/5).
Thus, our findings may not fully generalize to other tasks (e.g., multi-file projects, domain-specific code), languages, or future model architectures.
We mitigate this threat by covering diverse difficulty levels, 12 programming languages, and both open-source and proprietary models, and by reporting per-model and per-dataset results for transparent interpretation.

\subsection{Internal Validity}
Implementation issues (e.g., evaluation scripts, execution environment, decoding hyperparameters) and benchmark contamination could bias absolute Pass@1 scores.
We reuse or closely follow public evaluation pipelines where available, perform sanity checks on sampled tasks, and keep decoding settings consistent across paradigms within each model.
Since we do not control model pre-training corpora, we focus our conclusions on \emph{relative} differences between CoT paradigms under the same model, which are less sensitive to potential data leakage.

\subsection{Construct Validity}
We primarily use Pass@1 to measure code generation effectiveness and token consumption to approximate computational cost.
These metrics do not capture aspects such as readability, security, or runtime performance, and our information-theoretic analysis (via $I(Y;C|X)$) is conceptual rather than based on exact mutual information estimation.
Moreover, our mapping from concrete CoT implementations to abstract notions of "structured" and "reflective" reasoning is approximate; prompt and template design may influence behavior beyond paradigm structure, which we partially control in RQ4 but cannot fully disentangle.

\section{Conclusion and Future Work}
\label{sec:conclusion}

This study conducted a systematic empirical and information-theoretic study of Chain-of-Thought (CoT) paradigms for neural code generation.
Our results show that CoT effectiveness is jointly shaped by reasoning structure, model capacity, and task characteristics: structured paradigms (Self-Planning, SCoT) offer stable gains with much lower token cost, while reflective reasoning (Reasoning-CoT) provides marginally higher accuracy at substantially higher budget, and naive Zero-Shot CoT can even degrade performance.
We further find that smaller models and statically typed languages benefit more from high-quality structured reasoning, whereas larger models exhibit diminishing returns and dynamically typed languages are more compatible with reflective CoT, highlighting that both the \emph{quality} and \emph{design} of reasoning traces are critical.

Future work includes developing adaptive and hybrid CoT strategies that select or combine paradigms based on model scale, language, and task difficulty, and designing more direct measures of reasoning informativeness beyond Pass@1 and token statistics.
Extending our evaluation to multi-file, project-level, and tool-augmented code generation scenarios is another promising direction toward more robust and efficient code intelligence systems.

\section*{Declarations}

\noindent\textbf{Funding.} This work is supported by the Natural Science Foundation of Jiangsu Province under Grant No. BK20241194.  

\noindent\textbf{Ethical Approval.} Ethicalapprovalwas not required for this study.

\noindent\textbf{Informed Consent.} Informed consent was not applicablefor this study.

\noindent\textbf{Author Contributions.} 
\textbf{Naizhu Jin:} Data curation, Software, Writing - original draft.
\textbf{Zhong Li:} Conceptualization, Methodology, Writing - review \& editing, Supervision.
\textbf{Guang Yang:} Data curation, Software, Validation.
\textbf{Tian Zhang:} Writing - review \& editing, Validation.
\textbf{Qingkai Zeng:} Writing - review \& editing, Validation.

\noindent\textbf{Data Availability Statements.} Data available on request from the authors.

\noindent\textbf{Conflict of Interest.} The authors declare that they have no conflict of interest.

\noindent\textbf{Clinical Trial Number.} Cinical trial number: not applicable.

\bibliographystyle{spbasic}   
\bibliography{refs}           

@inproceedings{vaithilingam2022expectation,
  title={Expectation vs. experience: Evaluating the usability of code generation tools powered by large language models},
  author={Vaithilingam, Priyan and Zhang, Tianyi and Glassman, Elena L},
  booktitle={Chi conference on human factors in computing systems extended abstracts},
  pages={1--7},
  year={2022}
}

@article{yang2024important,
  title={How important are good method names in neural code generation? a model robustness perspective},
  author={Yang, Guang and Zhou, Yu and Yang, Wenhua and Yue, Tao and Chen, Xiang and Chen, Taolue},
  journal={ACM Transactions on Software Engineering and Methodology},
  volume={33},
  number={3},
  pages={1--35},
  year={2024},
  publisher={ACM New York, NY, USA}
}

@article{jiang2024survey,
  title={A survey on large language models for code generation},
  author={Jiang, Juyong and Wang, Fan and Shen, Jiasi and Kim, Sungju and Kim, Sunghun},
  journal={arXiv preprint arXiv:2406.00515},
  year={2024}
}

@article{kojima2022large,
  title={Large language models are zero-shot reasoners},
  author={Kojima, Takeshi and Gu, Shixiang Shane and Reid, Machel and Matsuo, Yutaka and Iwasawa, Yusuke},
  journal={Advances in neural information processing systems},
  volume={35},
  pages={22199--22213},
  year={2022}
}

@article{li2025structured,
  title={Structured chain-of-thought prompting for code generation},
  author={Li, Jia and Li, Ge and Li, Yongmin and Jin, Zhi},
  journal={ACM Transactions on Software Engineering and Methodology},
  volume={34},
  number={2},
  pages={1--23},
  year={2025},
  publisher={ACM New York, NY}
}

@article{roziere2023code,
  title={Code Llama: Open Foundation Models for Code},
  author={Roziere, Baptiste and Li, Raymond and Allal, Loubna Ben and others},
  journal={arXiv preprint arXiv:2308.12950},
  year={2023}
}

@article{chen2021evaluating,
  title={Evaluating large language models trained on code},
  author={Chen, Mark},
  journal={arXiv preprint arXiv:2107.03374},
  year={2021}
}

@inproceedings{peng2024humaneval,
  title={HumanEval-XL: A Multilingual Code Generation Benchmark for Cross-lingual Natural Language Generalization},
  author={Peng, Qiwei and Chai, Yekun and Li, Xuhong},
  booktitle={Proceedings of the 2024 Joint International Conference on Computational Linguistics, Language Resources and Evaluation (LREC-COLING 2024)},
  pages={8383--8394},
  year={2024}
}

@article{yang2024chain,
  title={Chain-of-thought in neural code generation: From and for lightweight language models},
  author={Yang, Guang and Zhou, Yu and Chen, Xiang and Zhang, Xiangyu and Zhuo, Terry Yue and Chen, Taolue},
  journal={IEEE Transactions on Software Engineering},
  year={2024},
  publisher={IEEE}
}

@article{austin2021program,
  title={Program synthesis with large language models},
  author={Austin, Jacob and Odena, Augustus and Nye, Maxwell and Bosma, Maarten and Michalewski, Henryk and Dohan, David and Jiang, Ellen and Cai, Carrie and Terry, Michael and Le, Quoc and others},
  journal={arXiv preprint arXiv:2108.07732},
  year={2021}
}

@inproceedings{wei2023towards,
  title={Towards greener yet powerful code generation via quantization: An empirical study},
  author={Wei, Xiaokai and Gonugondla, Sujan Kumar and Wang, Shiqi and Ahmad, Wasi and Ray, Baishakhi and Qian, Haifeng and Li, Xiaopeng and Kumar, Varun and Wang, Zijian and Tian, Yuchen and others},
  booktitle={Proceedings of the 31st ACM Joint European Software Engineering Conference and Symposium on the Foundations of Software Engineering},
  pages={224--236},
  year={2023}
}

@article{dai2024mhpp,
  title={Mhpp: Exploring the capabilities and limitations of language models beyond basic code generation},
  author={Dai, Jianbo and Lu, Jianqiao and Feng, Yunlong and Zeng, Guangtao and Ruan, Rongju and Cheng, Ming and Huang, Dong and Tan, Haochen and Guo, Zhijiang},
  journal={arXiv preprint arXiv:2405.11430},
  year={2024}
}

@article{zhuo2024bigcodebench,
  title={Bigcodebench: Benchmarking code generation with diverse function calls and complex instructions},
  author={Zhuo, Terry Yue and Vu, Minh Chien and Chim, Jenny and Hu, Han and Yu, Wenhao and Widyasari, Ratnadira and Yusuf, Imam Nur Bani and Zhan, Haolan and He, Junda and Paul, Indraneil and others},
  journal={arXiv preprint arXiv:2406.15877},
  year={2024}
}

@inproceedings{iyer2016summarizing,
  title={Summarizing source code using a neural attention model},
  author={Iyer, Srinivasan and Konstas, Ioannis and Cheung, Alvin and Zettlemoyer, Luke},
  booktitle={54th Annual Meeting of the Association for Computational Linguistics 2016},
  pages={2073--2083},
  year={2016},
  organization={Association for Computational Linguistics}
}

@inproceedings{yin2017syntactic,
  title={A Syntactic Neural Model for General-Purpose Code Generation},
  author={Yin, Pengcheng and Neubig, Graham},
  booktitle={Proceedings of the 55th Annual Meeting of the Association for Computational Linguistics (Volume 1: Long Papers)},
  year={2017},
  organization={Association for Computational Linguistics}
}

@inproceedings{feng2020codebert,
  title={CodeBERT: A Pre-Trained Model for Programming and Natural Languages},
  author={Feng, Zhangyin and Guo, Daya and Tang, Duyu and Duan, Nan and Feng, Xiaocheng and Gong, Ming and Shou, Linjun and Qin, Bing and Liu, Ting and Jiang, Daxin and others},
  booktitle={Findings of the Association for Computational Linguistics: EMNLP 2020},
  pages={1536--1547},
  year={2020}
}

@inproceedings{wang2021codet5,
  title={CodeT5: Identifier-aware Unified Pre-trained Encoder-Decoder Models for Code Understanding and Generation},
  author={Wang, Yue and Wang, Weishi and Joty, Shafiq and Hoi, Steven CH},
  booktitle={EMNLP 2021-2021 Conference on Empirical Methods in Natural Language Processing, Proceedings},
  pages={8696--8708},
  year={2021},
  organization={Association for Computational Linguistics (ACL)}
}

@article{lozhkov2024starcoder,
  title={Starcoder 2 and the stack v2: The next generation},
  author={Lozhkov, Anton and Li, Raymond and Allal, Loubna Ben and Cassano, Federico and Lamy-Poirier, Joel and Tazi, Nouamane and Tang, Ao and Pykhtar, Dmytro and Liu, Jiawei and Wei, Yuxiang and others},
  journal={arXiv preprint arXiv:2402.19173},
  year={2024}
}

@article{wei2022chain,
  title={Chain-of-thought prompting elicits reasoning in large language models},
  author={Wei, Jason and Wang, Xuezhi and Schuurmans, Dale and Bosma, Maarten and Xia, Fei and Chi, Ed and Le, Quoc V and Zhou, Denny and others},
  journal={Advances in neural information processing systems},
  volume={35},
  pages={24824--24837},
  year={2022}
}

@inproceedings{liurevisiting,
  title={Revisiting Chain-of-Thought in Code Generation: Do Language Models Need to Learn Reasoning before Coding?},
  author={Liu, Ren-Biao and Li, Anqi and Yang, Chaoding and Sun, Hui and Li, Ming},
  booktitle={Forty-second International Conference on Machine Learning},
  year={2025}
}

@article{miao2024chain,
  title={Chain of thought utilization in large language models and application in nephrology},
  author={Miao, Jing and Thongprayoon, Charat and Suppadungsuk, Supawadee and Krisanapan, Pajaree and Radhakrishnan, Yeshwanter and Cheungpasitporn, Wisit},
  journal={Medicina},
  volume={60},
  number={1},
  pages={148},
  year={2024},
  publisher={MDPI}
}

@article{jiang2024self,
  title={Self-planning code generation with large language models},
  author={Jiang, Xue and Dong, Yihong and Wang, Lecheng and Fang, Zheng and Shang, Qiwei and Li, Ge and Jin, Zhi and Jiao, Wenpin},
  journal={ACM Transactions on Software Engineering and Methodology},
  volume={33},
  number={7},
  pages={1--30},
  year={2024},
  publisher={ACM New York, NY}
}

@article{jin2025mscot,
  title={MSCoT: Structured Chain-of-Thought Generation for Multiple Programming Languages},
  author={Jin, Naizhu and Li, Zhong and Zhang, Tian and Zeng, Qingkai},
  journal={arXiv preprint arXiv:2504.10178},
  year={2025}
}

@article{nijkamp2022codegen,
  title={Codegen: An open large language model for code with multi-turn program synthesis},
  author={Nijkamp, Erik and Pang, Bo and Hayashi, Hiroaki and Tu, Lifu and Wang, Huan and Zhou, Yingbo and Savarese, Silvio and Xiong, Caiming},
  journal={arXiv preprint arXiv:2203.13474},
  year={2022}
}

@article{fried2022incoder,
  title={Incoder: A generative model for code infilling and synthesis},
  author={Fried, Daniel and Aghajanyan, Armen and Lin, Jessy and Wang, Sida and Wallace, Eric and Shi, Freda and Zhong, Ruiqi and Yih, Wen-tau and Zettlemoyer, Luke and Lewis, Mike},
  journal={arXiv preprint arXiv:2204.05999},
  year={2022}
}

@article{turpin2023language,
  title={Language models don't always say what they think: Unfaithful explanations in chain-of-thought prompting},
  author={Turpin, Miles and Michael, Julian and Perez, Ethan and Bowman, Samuel},
  journal={Advances in Neural Information Processing Systems},
  volume={36},
  pages={74952--74965},
  year={2023}
}

@article{yao2023tree,
  title={Tree of thoughts: Deliberate problem solving with large language models},
  author={Yao, Shunyu and Yu, Dian and Zhao, Jeffrey and Shafran, Izhak and Griffiths, Tom and Cao, Yuan and Narasimhan, Karthik},
  journal={Advances in neural information processing systems},
  volume={36},
  pages={11809--11822},
  year={2023}
}

@inproceedings{bi2024program,
  title={When do program-of-thought works for reasoning?},
  author={Bi, Zhen and Zhang, Ningyu and Jiang, Yinuo and Deng, Shumin and Zheng, Guozhou and Chen, Huajun},
  booktitle={Proceedings of the AAAI conference on artificial intelligence},
  volume={38},
  number={16},
  pages={17691--17699},
  year={2024}
}

@inproceedings{besta2024graph,
  title={Graph of thoughts: Solving elaborate problems with large language models},
  author={Besta, Maciej and Blach, Nils and Kubicek, Ales and Gerstenberger, Robert and Podstawski, Michal and Gianinazzi, Lukas and Gajda, Joanna and Lehmann, Tomasz and Niewiadomski, Hubert and Nyczyk, Piotr and others},
  booktitle={Proceedings of the AAAI conference on artificial intelligence},
  volume={38},
  number={16},
  pages={17682--17690},
  year={2024}
}

@article{fu2023chain,
  title={Chain-of-thought hub: A continuous effort to measure large language models' reasoning performance},
  author={Fu, Yao and Ou, Litu and Chen, Mingyu and Wan, Yuhao and Peng, Hao and Khot, Tushar},
  journal={arXiv preprint arXiv:2305.17306},
  year={2023}
}

@article{li2022competition,
  title={Competition-level code generation with alphacode},
  author={Li, Yujia and Choi, David and Chung, Junyoung and Kushman, Nate and Schrittwieser, Julian and Leblond, R{\'e}mi and Eccles, Tom and Keeling, James and Gimeno, Felix and Dal Lago, Agustin and others},
  journal={Science},
  volume={378},
  number={6624},
  pages={1092--1097},
  year={2022},
  publisher={American Association for the Advancement of Science}
}

@inproceedings{wang2023natural,
  title={Natural language to code: How far are we?},
  author={Wang, Shangwen and Geng, Mingyang and Lin, Bo and Sun, Zhensu and Wen, Ming and Liu, Yepang and Li, Li and Bissyand{\'e}, Tegawend{\'e} F and Mao, Xiaoguang},
  booktitle={Proceedings of the 31st ACM joint European software engineering conference and symposium on the foundations of software engineering},
  pages={375--387},
  year={2023}
}

@article{ma2023bridging,
  title={Bridging code semantic and llms: Semantic chain-of-thought prompting for code generation},
  author={Ma, Yingwei and Yu, Yue and Li, Shanshan and Jiang, Yu and Guo, Yong and Zhang, Yuanliang and Xie, Yutao and Liao, Xiangke},
  journal={arXiv preprint arXiv:2310.10698},
  year={2023}
}

@inproceedings{lyu2023faithful,
  title={Faithful chain-of-thought reasoning},
  author={Lyu, Qing and Havaldar, Shreya and Stein, Adam and Zhang, Li and Rao, Delip and Wong, Eric and Apidianaki, Marianna and Callison-Burch, Chris},
  booktitle={The 13th International Joint Conference on Natural Language Processing and the 3rd Conference of the Asia-Pacific Chapter of the Association for Computational Linguistics (IJCNLP-AACL 2023)},
  year={2023}
}

@inproceedings{jie2024interpretable,
  title={How interpretable are reasoning explanations from prompting large language models?},
  author={Jie, Yeo Wei and Satapathy, Ranjan and Goh, Rick and Cambria, Erik},
  booktitle={Findings of the Association for Computational Linguistics: NAACL 2024},
  pages={2148--2164},
  year={2024}
}

@article{stechly2024chain,
  title={Chain of thoughtlessness? an analysis of cot in planning},
  author={Stechly, Kaya and Valmeekam, Karthik and Kambhampati, Subbarao},
  journal={Advances in Neural Information Processing Systems},
  volume={37},
  pages={29106--29141},
  year={2024}
}

@inproceedings{diao2024active,
  title={Active prompting with chain-of-thought for large language models},
  author={Diao, Shizhe and Wang, Pengcheng and Lin, Yong and Pan, Rui and Liu, Xiang and Zhang, Tong},
  booktitle={Proceedings of the 62nd Annual Meeting of the Association for Computational Linguistics (Volume 1: Long Papers)},
  pages={1330--1350},
  year={2024}
}

@article{zhao2025trade,
  title={Trade-offs in large reasoning models: An empirical analysis of deliberative and adaptive reasoning over foundational capabilities},
  author={Zhao, Weixiang and Sui, Xingyu and Guo, Jiahe and Hu, Yulin and Deng, Yang and Zhao, Yanyan and Qin, Bing and Che, Wanxiang and Chua, Tat-Seng and Liu, Ting},
  journal={arXiv preprint arXiv:2503.17979},
  year={2025}
}

@article{zhao2025chain,
  title={Is chain-of-thought reasoning of llms a mirage? a data distribution lens},
  author={Zhao, Chengshuai and Tan, Zhen and Ma, Pingchuan and Li, Dawei and Jiang, Bohan and Wang, Yancheng and Yang, Yingzhen and Liu, Huan},
  journal={arXiv preprint arXiv:2508.01191},
  year={2025}
}
\end{document}